\documentclass[aps,pra,twocolumn,superscriptaddress]{revtex4-1}

\usepackage{float}
\usepackage{graphicx}% Include figure files
\usepackage{dcolumn}% Align table columns on decimal point
\usepackage{bm}% bold math
\usepackage{amsmath}
\usepackage{amssymb}
\usepackage{latexsym}
\usepackage{epsfig}
\usepackage{amsbsy}
\usepackage{array}
\usepackage{bm}
\usepackage{comment}
\usepackage{makecell}
\usepackage{mathrsfs}
\usepackage{xcolor}
\usepackage{ulem}
\usepackage[colorlinks=true,citecolor=blue,urlcolor=blue,linkcolor=blue]{hyperref}

\makeatletter
\newcommand{\rmnum}[1]{\romannumeral #1}
\newcommand{\Rmnum}[1]{\expandafter\@slowromancap\romannumeral #1@}
\makeatother

\newcommand{\ac}[0]{{a_{\rm c}}}
\newcommand{\omegac}[0]{{\omega_{\rm c}}}
\newcommand{\ud}[0]{{\rm d}}

\newcommand{\dw}[0]{D(\omega_{\rm c})}

\newcommand{\phimn}[1]{{\Delta \phi_{ #1}}}
\newcommand{\pwpx}[3]{{\frac{\partial  \widetilde{\omega}_#1(t_#3)  }{\partial x_#2}}}
\newcommand{\pwpxt}[2]{{\frac{ \partial \widetilde{\omega}_#1(t)  }{\partial x_#2}}}
\newcommand{\tg}[0]{{ T_{\rm g}}}
\newcommand{\dx}[2]{{\delta x_#1(t_#2)}}
\newcommand{\dxt}[1]{{\delta x_#1(t)}}

\newcommand{\gammacorr}[2]{{\Gamma}^#1_{#2,\rm corr}}
\newcommand{\gammawhite}[2]{{\Gamma}^#1_{#2,\rm white}}
\newcommand{\gammaf}[2]{{\Gamma}^#1_{#2,1/f}}

\begin{document}

\title{Coupler-Assisted Controlled-Phase Gate with Enhanced Adiabaticity}

\author{Ji Chu}
\affiliation{Shenzhen Institute for Quantum Science and Engineering, Southern University of Science and Technology, Shenzhen, Guangdong, China}

\author{Fei Yan}
\email{yanf7@sustech.edu.cn}
\affiliation{Shenzhen Institute for Quantum Science and Engineering, Southern University of Science and Technology, Shenzhen, Guangdong, China}
\affiliation{Guangdong Provincial Key Laboratory of Quantum Science and Engineering,
	Southern University of Science and Technology, Shenzhen, Guangdong, China}
\affiliation{Shenzhen Key Laboratory of Quantum Science and Engineering,
	Southern University of Science and Technology, Shenzhen, Guangdong, China}

\begin{abstract}	
{
High-fidelity two-qubit entangling gates are essential building blocks for fault-tolerant quantum computers.
Over the past decade, tremendous efforts have been made to develop scalable high-fidelity two-qubit gates with superconducting quantum circuits.
Recently, an easy-to-scale controlled-phase gate scheme that utilizes the tunable-coupling architecture with fixed-frequency qubits [Phys.\ Rev.\ Lett.\ \textbf{125}, 240502; Phys.\ Rev.\ Lett.\ \textbf{125}, 240503] has been demonstrated with high fidelity and attracted broad interest.
However, in-depth understanding of the underlying mechanism is still missing, preventing us from fully exploiting its potential.
Here we present a comprehensive theoretical study, explaining the origin of the high-contrast ZZ interaction. Based on improved understanding, we develop a general yet convenient method for shaping an adiabatic pulse in a multilevel system, and identify how to optimize the gate performance from design.
Given state-of-the-art coherence properties, we expect the scheme to potentially achieve a two-qubit gate error rate near $10^{-5}$, which would drastically speed up the progress towards fault-tolerant quantum computation.
}
\end{abstract}

\maketitle
	
\section{Introduction}

Fault-tolerant quantum computing relies on high-fidelity quantum operations.
Currently, two-qubit quantum logic gates are the performance bottleneck in various platforms~\cite{arute2019quantum,watson2018programmable,figgatt2019parallel,levine2019parallel}, because it is extremely challenging to tailor qubit interactions for precise state manipulation, especially when faster gates are favored for suppressing decoherence errors.
In addition, increasing the system size exacerbates the situation with added complexity.
Therefore, a high-fidelity yet easy-to-scale two-qubit
gate scheme is the key to scalable quantum information processing.

During the past decade, extensive efforts have been made to achieve high-speed, high-fidelity, and robust two-qubit gates with superconducting quantum circuits~\cite{krantz2019quantum,blais2021circuit}.
In the conventional architecture, qubits are usually coupled directly or via a resonator bus~\cite{majer2007coupling}. Various two-qubit gate schemes - such as adiabatic gates~\cite{dicarlo2009demonstration}, diabatic gates~\cite{neeley2010generation,dewes2012characterization}, microwave gates~\cite{chow2011simple,poletto2012entanglement,chow2013microwaveactivated,caldwell2018parametrically}, resonator-induced gates~\cite{paik2016experimental} - have been proposed and demonstrated.
Although some works have shown gate error rate less than 1\%~\cite{barends2014superconducting,sheldon2016procedure,kjaergaard2020programming,li2019realisation,barends2019diabatic,rol2019fast,hong2020demonstration,negirneac2020high}, performance of these schemes are fundamentally limited by the low ON/OFF ratio in the interaction strength and by the frequency crowding issue, both due to the presence of the always-on coupling~\cite{brink2018device}.
These issues are successfully addressed by introducing a tunable coupler~\cite{chen2014qubit}, which is an independent control knob over the coupling strength.
This new control freedom enables rapid switch of qubit interaction and isolation of target qubits from their neighbors.
Based on a trending tunable-coupling architecture~\cite{yan2018tunable}, diabatic two-qubit gates have been demonstrated with high speed and high fidelity~\cite{arute2019quantum, foxen2020demonstrating,sung2020realization}.
However, the scheme requires tunability for both the qubits and the coupler, leading to a more complex control setup and cumbersome calibration procedures.

Recently, a simpler and easier-to-scale controlled-phase gate scheme based on coupler-assisted ZZ interaction with fixed-frequency qubits has been proposed and demonstrated by the authors and other researchers~\cite{collodo2020implementation,xu2020high}.
Unlike microwave gates performed with fixed-frequency qubits (typically a-few-hundred-nanosecond long)~\cite{mckay2016universal,mundada2019suppression,kandala2020demonstration,cai2021perturbation,sete2021parametric}, the adiabatic scheme can be fast yet precise, achieving 99.5\% gate fidelity mostly limited by $T_1$ in a 30~ns gate~\cite{xu2020high}.
Other works related to the scheme have also been shown recently~\cite{stehlik2021tunable,xu2021realization,jin2021implementing}.
Although the scheme is promising, there still a lack of a well-rounded and in-depth understanding of its physical mechanism. This prevents us from further exploiting the scheme for better performance.

In this work, we present a comprehensive theoretical study of the coupler-assisted adiabatic controlled-phase gate scheme.
Strong ZZ interaction originates from nontrivial level repulsion between double-excitation states. To access such interaction, an intermediate state - assisted by the tunable coupler - has to be used.
The condition of suppressing residual ZZ interaction is identified in various circuit configurations.
To optimize the gate performance from the control perspective, we propose a generally applicable and practically convenient method for shaping an adiabatic pulse in a multilevel system, promising low nonadiabatic errors and high robustness against parameter uncertainty and pulse distortion.
From the design perspective, it is highly desirable to have large qubit-qubit detuning.
By investigating various types of decoherence errors, we find that the $T_1$ process and pure dephasing are currently the leading error sources.
Finally, pros and cons of different extension schemes are discussed.

The outline of this manuscript is as follows.
In Sec.~\Rmnum{2}, we introduce the coupler-assisted system, reveal the underlying mechanism of strong ZZ interaction, and study residual ZZ interaction.
In Sec.~\Rmnum{3}, we introduce the new method for adiabatic pulse engineering, show how to optimize circuit parameters, and discuss the effect of stray coupling.
In Sec.~\Rmnum{4}, we investigate how various types of noise and decoherence processes affect the gate performance based on a revisited error model connecting to two-qubit randomized benchmarking (RB) experiment.
We compare different extension schemes in Sec.~\Rmnum{5} and conclude in Sec.~\Rmnum{6}.

\section{ Tunable ZZ interaction}
	
\subsection{Circuit implementation and definitions}

\begin{figure}
	\includegraphics[width=8.5cm]{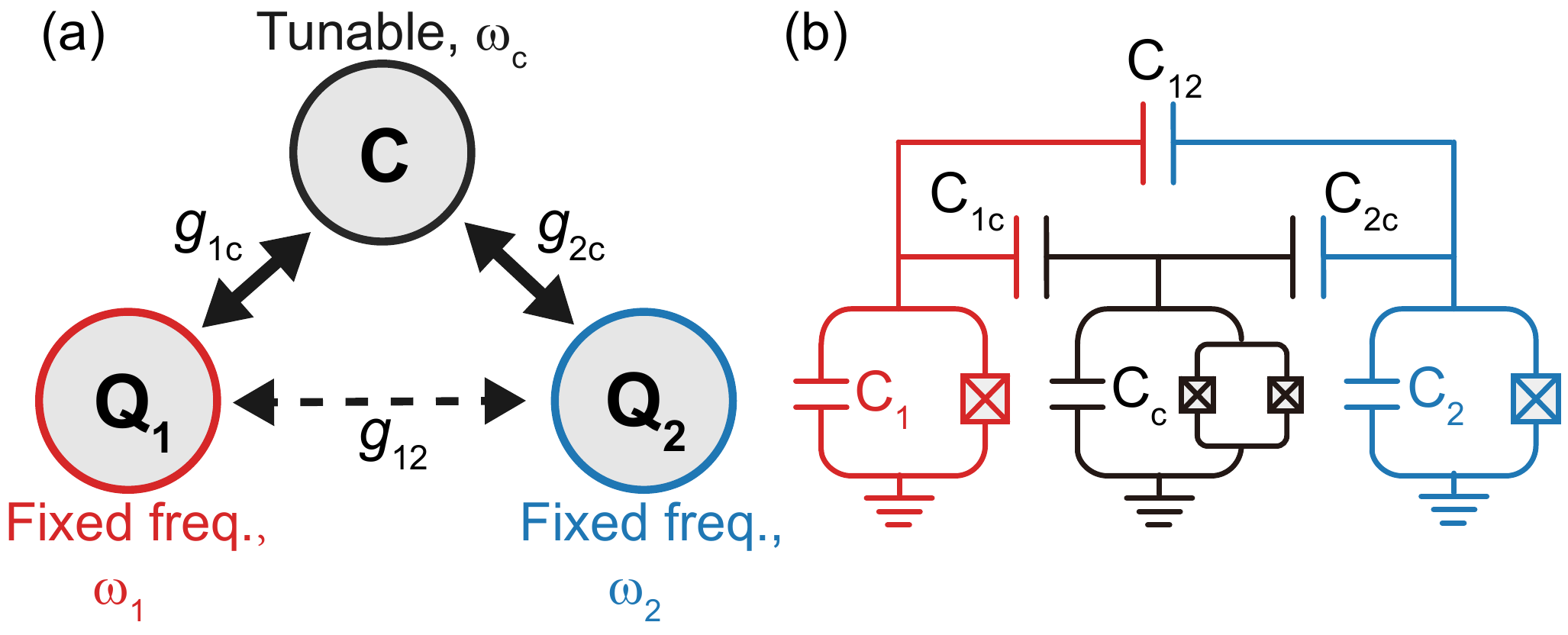}
	\caption{ 
		(a) Conceptual sketch of the tunable-coupling architecture, where the direct coupling strength between two fixed-frequency qubits $\rm Q_1$ and $\rm Q_2$ ($g_{12}$) is considered much weaker than the coupling between the qubits and the coupler C ($g_{\rm 1c}$, $g_{\rm 2c}$).
		(b) Circuit diagram for implementing the architecture described in (a) with supercondcuting quantum circuits. Single-junction transmon qubits and a split-transmon coupler are all capacitively coupled to each other. Both the qubit-coupler coupling capacitances $C_{i\rm c},(i=1,2)$ and the qubit-qubit coupling capacitance $C_{12}$ are much smaller than the shunt capacitances of the transmons $C_i (i=1,2,\rm c)$.
		\label{fig:FIG 1}}
\end{figure} 

In a generic model, the tunable-coupling architecture consists of three modes with exchange interactions.
As sketched in Fig.~\ref{fig:FIG 1}(a), two qubits ($\rm Q_1$ and $\rm Q_2$) couple to a coupler (C) with coupling strength $g_{\rm 1c}$ and $g_{\rm 2c}$, as well as to each other with a weaker coupling strength $g_{12}$.
The coupler frequency is tunable, while the qubit frequencies are fixed.
Each mode is an anharmonic oscillator with frequency $\omega_{i}$ and anharmonicity $\alpha_i$ ($i=1,2,\rm c$). Without loss of generality, we assume $\omega_1\geqslant\omega_2$. 
The full system Hamiltonian can be written as ($\hbar=1$)
\begin{equation}
\begin{aligned}
H & = H_0 + H_{\rm int}, \quad \rm where\\
H_{0} &= \sum_{i=1,2,\rm c}(\omega_{i}a_i^{\dagger}a_i+\frac{\alpha_i}{2} a_i^{\dagger}a_i^{\dagger}a_ia_i) \quad \rm and \\
H_{\rm int} &= \sum_{i=1,2} g_{i \rm c}( a_i^{\dagger}+a_i )( a_{\rm c}^{\dagger}+a_{\rm c} ) 
+g_{12}( a_1^{\dagger}+a_1 )( a_2^{\dagger}+a_2 ). \label{eq:hamiltonian_full}
\end{aligned} 
\end{equation}
Here, $H_0$ is the uncoupled or bare Hamiltonian, $H_{\rm int}$ is the interaction Hamiltonian, $a_i$($a_i^{\dagger}$) is the annihilation(creation) operator of the corresponding mode. 

The architecture can be conveniently implemented with superconducting quantum circuits and, more importantly, the transmon qubits~\cite{koch2007charge}, currently the leading technology for building scalable quantum information processors.
A typical circuit implementation is illustrated in Fig.~\ref{fig:FIG 1}(b).
Here, fixed-frequency qubits are realized with the more coherent single-junction transmon qubits; the tunable coupler is a two-junction transmon qubit, the frequency of which is tunable by external magnetic flux threading the SQUID loop.
Note that the coupler frequency $\omega_{\rm c}$ is the only tunable parameter in this circuit.
The exchange couplings (capacitive) have frequency-dependent coupling strength, $g_{ij} = \rho_{ij} \sqrt{\omega_{i}\omega_{j}}$, where $\rho_{i\rm c} = \frac{C_{i\rm c}}{2\sqrt{C_{i} C_{\rm c}}}$ and $\rho_{12} = \frac{C_{\rm 1c}C_{\rm 2c}/C_{\rm c} +C_{12}}{2\sqrt{C_1 C_2}}$ are constants depending only on the circuit geometry.
Using Schrieffer-Wolff transformation, one can derive the effective exchange-coupling strength between qubits,
\begin{equation}
\begin{aligned}
\widetilde{g} = g_{12}+\frac{1}{2}g_{\rm 1c}g_{\rm 2c}(\frac{1}{\Delta_{\rm 1c}}+\frac{1}{\Delta_{\rm 2c}}-\frac{1}{\Sigma_{\rm 1c}}-\frac{1}{\Sigma_{\rm 2c}}), \label{eq:g_effective}
\end{aligned} 
\end{equation}
where $\Delta_{ij}=\omega_{i}-\omega_{j}$ and $\Sigma_{ij}=\omega_{i}+\omega_{j}$.
It can be seen that the effective coupling $\widetilde{g}$ is the net result of the direct qubit-qubit exchange coupling and the indirect exchange via the coupler~\cite{yan2018tunable}. 

\begin{figure}
	\includegraphics[width=8.5cm]{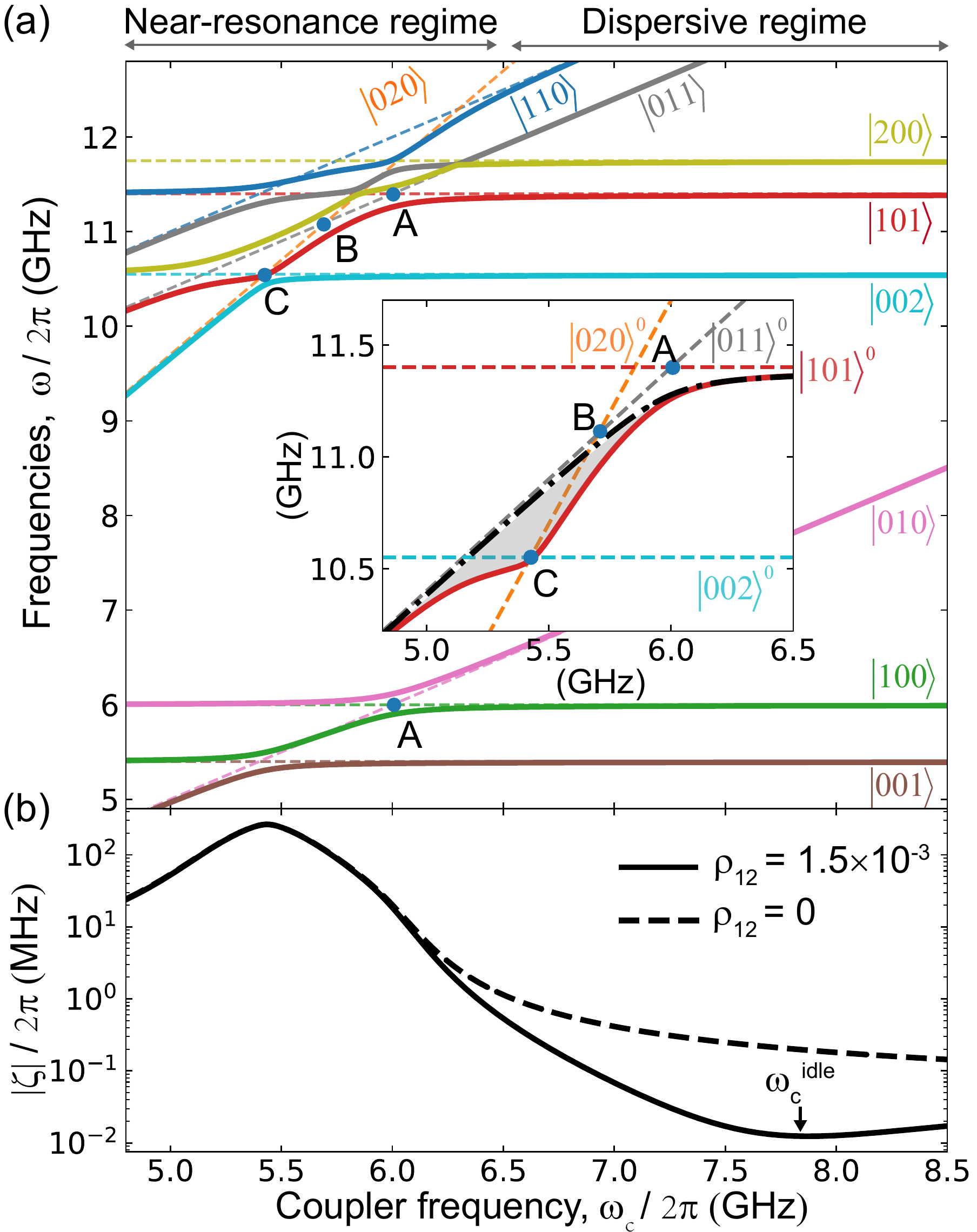}
	\caption{ (a) Energy-level spectra of the circuit Hamiltonian in Eq.~(\ref{eq:hamiltonian_full}) as a function of the coupler frequency $\omega_{\rm c}/2\pi$.
	Solid (dashed) lines correspond to energies (frequencies) of adiabatic (diabatic) states.
	Point A, B and C indicate level crossings between different pairs of diabatic states.
	The horizontal range is divided into the near-resonance regime and the dispersive regime.
	Circuit parameters are: $\omega_{1,2} / 2\pi = 6.0, 5.4$~GHz, $\alpha_{1,2,{\rm c}} / 2\pi = -250, -250, -300$~MHz, $\rho_{\rm 1c,{\rm 2c},12} = 0.018,0.018,0.0015$.
	Inset is the zoom-in of the interested region with avoided crossings.
	The dotted black line denotes the sum frequency of $|100\rangle$ and $|001\rangle$, so the difference between the dotted black line and the solid red line ($|101\rangle$) - the shaded region - indicates the ZZ interaction strength $\zeta$.
	(b) Absolute value of $\zeta/2\pi$ versus $\omega_{\rm c}/2\pi$ with (solid) and without (dashed) direct qubit-qubit coupling. 
	\label{fig:FIG 2}}
\end{figure}

Since $H_0$ is diagonal, its eigenstate - the bare state or diabatic state - can be denoted as $|n_1n_{\rm c}n_2\rangle^0 = |n_1\rangle \otimes |n_{\rm c}\rangle \otimes |n_2\rangle$, where $|n_{i}\rangle$ ($n_{i}=0,1,2...$) is the photon number (Fock) state of the individual mode. 
Their sum gives the total excitation number $n_{\rm ex}=n_1+n_{\rm c}+n_2$.
The energy spectra (versus $\omega_{\rm c}$) of the diabatic states are shown as the dashed lines in Fig.~\ref{fig:FIG 2}(a). 
The eigenstate of the full Hamiltonian $H$ - the dressed state or adiabatic state $|n_1n_{\rm c}n_2\rangle$ - is, however, superposition of the diabatic states in general. 
Since the couplings are transverse ($[H_{\rm int},H_0]\neq0$), crossings between interacting levels are avoided. 
For convenience, we label the adiabatic states by (ascending) order of their eigenvalues, such that, as shown as solid lines in Fig.~\ref{fig:FIG 2}(a), the level of an adiabatic state is continuous across the interested range. For example, the adiabatic state $|101\rangle$ is shown as the solid red line which has no intersection with other lines.
In the label of $|n_1n_{\rm c}n_2\rangle$, the photon numbers are chosen to be the same as in its closest diabatic state $|n_1n_{\rm c}n_2\rangle^0$ in the \textit{dispersive} limit, a regime where modes are far detuned ($\Delta_{ij} \gg g_{ij}$) and hence the diabatic state and the adiabatic state are almost fully overlaped ($\left| \langle n_1n_{\rm c}n_2 | n_1n_{\rm c}n_2\rangle^0 \right| \approx 1$).

Under our definition, in the near-resonance regime where crossings are forming between diabatic states, an adiabatic state $|n_1n_{\rm c}n_2\rangle$ may, at different $\omega_{\rm c}$, have major overlap with different diabatic states instead of $|n_1n_{\rm c}n_2\rangle^0$.
For example, as shown in Fig.~\ref{fig:FIG 2}(a), point A indicates the crossing between $|101\rangle^0$ and $|011\rangle^0$; point B indicates the crossing between $|011\rangle^0$ and $|020\rangle^0$; point C indicates the crossing between $|020\rangle^0$ and $|002\rangle^0$.
At these bias points, the adiabatic state is approximately an equal superposition of interacting diabatic states.
Therefore, the $|101\rangle$ state (solid red line) has more overlap with $|101\rangle^0$ (dashed red line) to the right side of crossing A; more overlap with $|011\rangle^0$ (dashed gray line) between crossings A and B; more overlap with $|020\rangle^0$ (dashed orange line) between crossings B and C.
This is also evidenced by the asymptotic behavior between the solid red line and each dashed line in each region.

Next, we clarify a few definitions.
The full Hilbert space $\mathcal{H}$ has infinite dimensions, but only low-energy states are of physical interest.
We define $\mathcal{M}_{n_{\rm ex}}$ as the manifold or energy band having a total excitation number $n_{\rm ex}$. For example, $\mathcal{M}_1 = \{|100\rangle,|010\rangle,|001\rangle\}$ is the single-excitation manifold; $\mathcal{M}_2 = \{|101\rangle,|110\rangle,|011\rangle,|200\rangle,|020\rangle,|002\rangle\}$ is the double-excitation manifold.
We also define $\mathcal{S} = \{|000\rangle,|001\rangle,|100\rangle,|101\rangle\}$ as the (two-qubit) computational subspace for logic operations. Note that the computational bases are the \textit{adiabatic} states. It is usually convenient to choose energy eigenstates at the idling bias or bias where to perform single-qubit operations as computational basis states.
Accordingly, the leakage subspace $\mathcal{L} = \overline{\mathcal{S}}$ is the complement of computational subspace.

\subsection{Coupler-assisted ZZ interactions}

We may rewrite the full Hamiltonian $H$ in its instantaneous energy eigenbases - that is, the adiabatic states which are $\omega_{\rm c}$-dependent - and truncate it to the computational subspace,
\begin{equation}
\begin{aligned}
\widetilde{H} &= \widetilde{\omega}_{1}|100\rangle \langle 100| + \widetilde{\omega}_{2}|001\rangle \langle 001| \\
&+ (\widetilde{\omega}_{1}+\widetilde{\omega}_{2}+\zeta )|101\rangle \langle 101|,
\label{eq:hamiltonian_subspace}
\end{aligned}
\end{equation}
where $\widetilde{\omega}_i=\widetilde{\omega}_i(\omega_{\rm c})$ ($i=1,2$) are the eigen-energies of $|100\rangle$ and $|001\rangle$ respectively, $\zeta=\zeta(\omega_{\rm c})$ is, by our definition, the ZZ interaction strength. All these energies or frequencies are parameterized by $\omega_{\rm c}$. These three states are shown as solid lines (green, brown, red) in Fig.~\ref{fig:FIG 2}(a). The energy of the ground state $|000\rangle$ is approximately 0 across the range.

The ZZ interaction $\zeta$ is a non-trivial term which generates entangled phase, and thus is at the center of our interest.
By comparing the sum frequency of $|100\rangle$ and $|001\rangle$ (dotted black line in Fig.~\ref{fig:FIG 2}(a), inset) to that of $|101\rangle$ (solid red line), we find that their difference - $\zeta$ by definition - becomes drastically greater, as we adjust the coupler frequency from the dispersive regime to the near-resonance regime (from right to left as in the figure).
Figure \ref{fig:FIG 2}(b) shows that the magnitude of $\zeta(\omega_{\rm c})$ changes continuously from $\sim$10~kHz to above 100~MHz. It is essentially a coupling switch with a high ON/OFF ratio ($>10^4$) and with only single control knob, the coupler frequency $\omega_{\rm c}$.
If adiabatically adjusting $\omega_{\rm c}$ from a low-$|\zeta|$ (idling) bias in the dispersive regime to the high-$\zeta$ regime and back to idling, one obtains a controlled-phase gate as demonstrated in previous experiments~\cite{xu2020high,collodo2020implementation}.

To understand the origin of $\zeta$, suppose that the qubits and the coupler are all two-level systems. Hence, all the two-photon ($n_i=2$) states, $|200\rangle^0$, $|020\rangle^0$ and $|002\rangle^0$, would be absent. Note that, by our definition, a double-excition state may not be a two-photon state.
Since the level-repulsion effect between $|100\rangle^0$ and $|010\rangle^0$ is symmetric with that between $|101\rangle^0$ and $|011\rangle^0$, the adiabatic state $|101\rangle$ (solid red line) would exactly follow the dotted black line, leading to zero $\zeta$.
The non-trivial ZZ interaction must have arised from interactions with these two-photon states (see Appendix A for more discussions).

To better explain how $\zeta$ rises, we may track the system dynamics - in particular, the $|101\rangle$ state - by following an adiabatical sweep of the coupler from an idling bias into the near-resonance regime. The process is analyzed step-by-step as follows.

(\rmnum{1}) In the dispersive regime, the adiabatic state $|101\rangle$ has near-unity overlap with $|101\rangle^0$, resulting in small $|\zeta|$ (below $1$ MHz), mostly from level repulsion against $|200\rangle^0$ in the case shown.

(\rmnum{2}) As we pass the first avoided crossing A, the $|101\rangle$ level is bent due to strong $|101\rangle$-$|011\rangle$ coupling with strength $g_{1 \rm c}$. However, $|\zeta|$ is still insignificant because the bending effect is largely cancelled by a similar bending of the $|100\rangle$ level (solid green line). 

(\rmnum{3}) As we pass crossing A and move towards crossing B, the adiabatic state $|101\rangle$ is now mainly overlapped with the diabatic state $|011\rangle^0$. 
Note that $|011\rangle^0$ is also strongly coupled to $|020\rangle^0$ with strength $\sqrt{2}g_{2 \rm c}$.
Therefore, when moving closer to the second avoided crossing B, the $|101\rangle$ level gets bent again.
Meanwhile, there is no such effect on the $|100\rangle$ line anymore.
The non-trivial $|011\rangle^0$-$|020\rangle^0$ crossing leads to drastic rise in $|\zeta|$.

(\rmnum{4}) After passing over crossing B, the adiabatic state $|101\rangle$ becomes mainly overlapped with $|020\rangle^0$.
The gap between the $|101\rangle$ line and dotted black line continues to widen due to the difference in the spectral slopes.
$|\zeta|$ increases almost at a near-unity rate, i.e.\ $\rm d|\zeta|/\rm d\omega_{\rm c} \sim 1$.

(\rmnum{5}) Crossing C is an avoided crossing with a much smaller gap caused by the weak second-order coupling between $|020\rangle^0$ and $|002\rangle^0$.
In practice, we prefer to stay away from small avoided crossings which are harmful to adiabatic condition.
In the adiabatic controlled-phase gate, a reasonable pulse is expected to sweep through crossings A and B but stops far before crossing C, so as to maximize ZZ interaction while avoiding non-adiabatic effect. 

From above, the picture is now clear. The nontrivial level repulsion between $|020\rangle^0$ and $|011\rangle^0$ is the origin of strong ZZ interaction in this configuration, and the $|011\rangle^0$ state plays the role of an intermediate state that opens up the path to the nontrivial interaction.
Note that $|200\rangle^0$ and $|002\rangle^0$ are also possible origins of ZZ interaction in certain configurations.
Due to strong qubit-coupler coupling, the avoided crossings A and B both have large gaps (>200~MHz), leading to enhanced adiabaticity when a pulse sweeps through them.
Therefore, to enable this scheme, it is crucial to engineer the spectral structure so that, when adiabatically adjusting the coupler frequency, the $|101\rangle$ state would first evolve into an intermediate state - $|011\rangle^0$ in this case - which can directly interact with a two-photon state such as $|020\rangle^0$.
In our example, it is necessary to use \textit{weakly anharmonic} systems such as transmon qubits as the tunable coupler in order to ensure the process to proceed in the correct order and to be free from small-gap crossings.
From this perspective, our scheme is an ideal fit for transmon qubits, which are known for high reproducibility but often criticized for weak anharmonicity.

\subsection{Residual ZZ in the dispersive regime}

\begin{figure}
	\includegraphics[width=8.5cm]{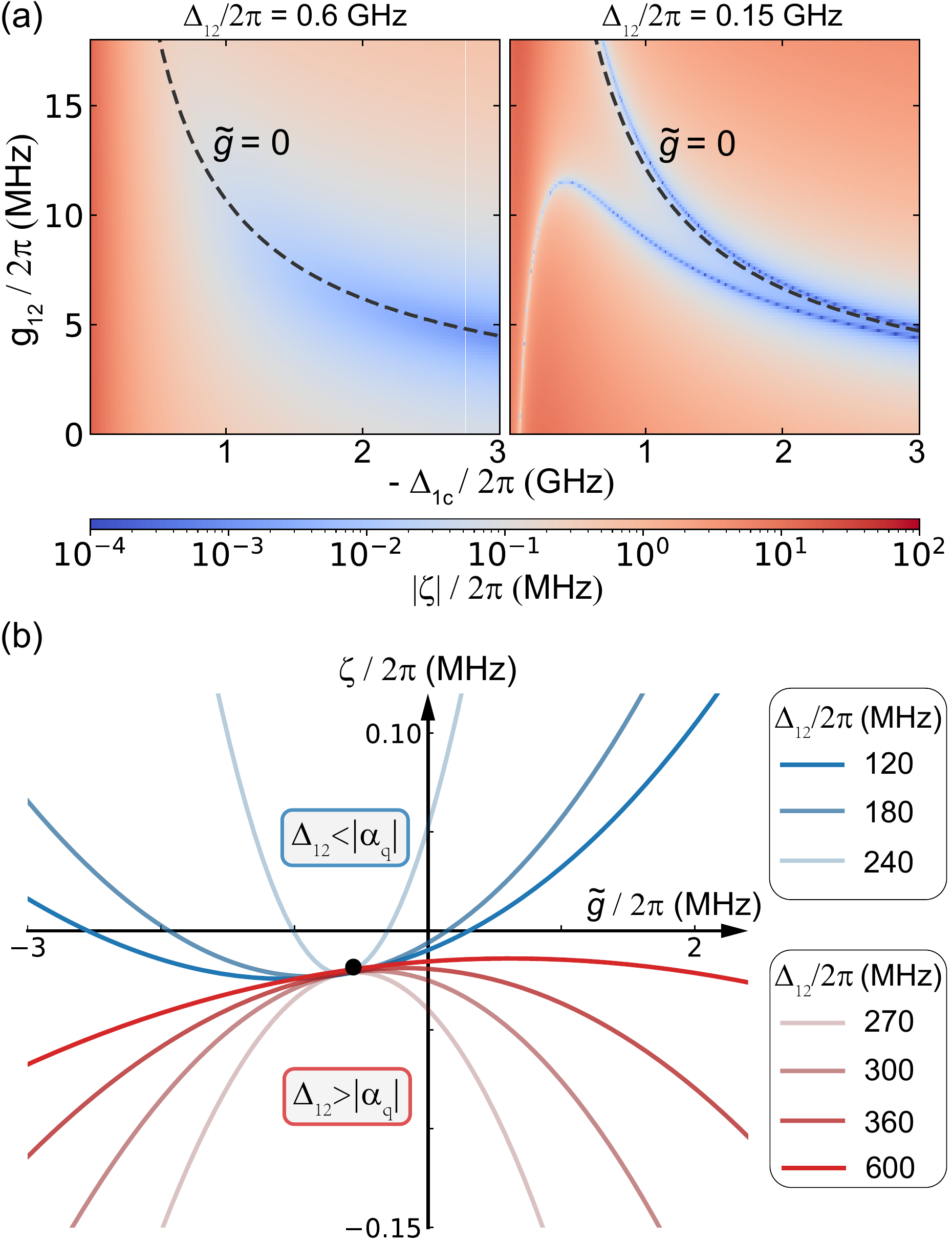}
	\caption{
		(a) Absolute value of ZZ interaction strength $|\zeta|$ versus qubit-coupler detuning $\Delta_{\rm 1c}$ and direct qubit-qubit coupling strength $g_{12}$, given qubit-qubit detuning $\Delta_{12} / 2\pi = 600$ MHz (left panel) and 150 MHz (right panel).
		The coupling strength between the qubits and the coupler are $g_{\rm 1c,\rm 2c}/2\pi = 120, 100$ MHz and the anharmonicities are $\alpha_{1,2,\rm c} / 2\pi = -250, -250, -300$ MHz. The dashed black lines indicate zero effective qubit-qubit coupling, i.e.\ $\widetilde{g} = 0$.
		(b) ZZ interaction strength $\zeta$ (numerically simulated) versus effective qubit-qubit coupling strength $\widetilde{g}$ for different qubit-qubit detunings $\Delta_{12}$. 
        The direct qubit-qubit coupling strength $g_{12} / 2\pi =8$ MHz.
        According to Eq.~(\ref{eq:zeta_quad}), $\zeta$ is approximately a quadratic function of $\widetilde{g}$. The parabolas open upward in the small-detuning regime ($\Delta_{12}<|\alpha_{\rm q}|$) or downward in the large-detuning regime ($\Delta_{12}<|\alpha_{\rm q}|$). These parabolas share a common point (black dot) at $\widetilde{g} = \alpha_{\rm q}\nu$ and $\zeta = (8\alpha_{\rm c}+4\alpha_{\rm q})\nu^2$. 
		} \label{fig:FIG 3}
\end{figure} 

A typical quantum circuit is constructed by modular single-qubit, two-qubit, and idling operations.
The residual ZZ interaction during single-qubit and idle operations - often manifested as a kind of coherent crosstalk~\cite{mundada2019suppression} - compromises gate fidelity and hinders the realization of fault-tolerant quantum computation~\cite{gambetta2012characterization,andersen2020repeated,huang2020alibaba}.
Here, we show how to improve ZZ suppression by properly choosing the circuit design parameters.

Recent development has focused on two approaches for suppressing residual ZZ interaction. One is to directly couple (no tunable coupler needed) two qubits with opposite anharmonicity~\cite{zhao2020high,ku2020suppression}.
The other approach is to take advantage of the tunable-coupling design, which is compatible with all types of qubits including the transmon qubits.
This work focuses on the latter.
As shown in Fig.~\ref{fig:FIG 2}(b), with or without a direct coupling, it has little influence on ZZ interaction in the near-resonance regime, but makes a big difference in the dispersive regime. The residual coupling is more than an order of magnitude higher without a direct coupling.

Figure 3(a) plots $|\zeta|$ as a function of the qubit-coupler detuning $\Delta_{\rm 1c}$ and the direct coupling strength $g_{12}$, for both the large-detuning ($\Delta_{12}>|\alpha_{\rm q}|$, where $\alpha_{\rm q}=\alpha_{\rm 1,2}$, left panel) and small-detuning case ($\Delta_{12}<|\alpha_{\rm q}|$, right panel). The black dashed line indicates the condition of zero effective (net) qubit-qubit coupling ($\widetilde{g}=0$) which, as can be seen, almost agrees to that of minimum residual ZZ interaction in both cases.
This confirms that by introducing an additional coupling path - the direct qubit-qubit coupling - to zero out $\widetilde{g}$, residual ZZ interaction can be substantially suppressed according to the same condition regardless of how much the qubits are detuned.

In the tunable-coupling architecture, the dispersive ZZ interaction strength $\zeta$ can be calculated using fourth-order perturbation theory (see full result in Appendix B).
Given that $\Sigma_{i \rm c} \gg |\Delta_{i \rm c}| \gg g_{i \rm c} \gg g_{12}$ in realistic situations, the result is simplified to
\begin{equation}
\begin{aligned}
\zeta & \approx 
\frac{2 \left[ (\alpha_{1}+\alpha_{2})\widetilde{g}^2 - 2\nu(2\alpha_{1}\alpha_{2} + (\alpha_{1}-\alpha_{2})\Delta_{12} )\widetilde{g}  \right] }{(\Delta_{12}+\alpha_{1})(\Delta_{12}-\alpha_{2})} \\
& + 2\nu^2 \left[ 4\alpha_{\rm c} + \frac{(\alpha_{1}+\alpha_{2})\Delta_{12}^2}{(\Delta_{12}+\alpha_{1})(\Delta_{12}-\alpha_{2})} \right] \\
& \overset{ \alpha_{1}=\alpha_{2} = \alpha_{\rm q} }{\longrightarrow} = \frac{4\alpha_{\rm q}}{\Delta_{12}^2 - \alpha_{\rm q}^2} (\widetilde{g} - \alpha_{\rm q} \nu )^2 + 4(2\alpha_{\rm c}+\alpha_{\rm q})\nu^2,
\label{eq:zeta_quad}
\end{aligned} 
\end{equation}
where $\nu = g_{\rm 1c}g_{\rm 2c}/(2\Delta_{\rm 1c}\Delta_{\rm 2c})$ is a small ($\sim \rm 10^{-3}$) dimensionless quantity in the dispersive limit.

There are a few implications from Eq.~(\ref{eq:zeta_quad}).
In the current representation, $\zeta$ is a quadratic function of $\widetilde{g}$.
As in the first line of the equation, the dominant $\widetilde{g}^2$ term can be cancelled by using qubits with opposite anharmonicity with $\alpha_1=-\alpha_2$. This explains why it is possible to eliminate residual ZZ interaction by using, for example, a positively anharmonic qubit such as capacitively shunted flux qubit~\cite{yan2016flux} together with a transmon qubit.

The third line of Eq.~(\ref{eq:zeta_quad}) corresponds to the case of an all-transmon circuit, in which both qubits have similar and negative anharmonicity.
The quadratic relation - the parabola shown in Fig.~\ref{fig:FIG 3}(b) - are categorized into the small-detuning (upward opening) and large-detuning (downward opening) regime.
It is possible to achieve zero $\zeta$ in the small-detuning case with a proper choice of $\widetilde{g}$.
For large detuning, it is impossible for $\zeta$ to be exactly zero, but the optimal residual ZZ interaction, $\zeta \propto \nu^4 = (g_{\rm ic}/\Delta_{\rm ic})^4$, can still be made small enough ($\sim$10~kHz).
To compare, in the coupler-free design~\cite{barends2014superconducting}, $\widetilde{g} \equiv g_{\rm qq}$ and $|\zeta|\propto g_{\rm qq}^2 / \Delta_{\rm qq}$.
For a reasonably small residual ZZ interaction ($|\zeta^{\rm idle}|/2\pi \sim 100$ kHz), $g_{\rm qq}$ is limited to tens of megahertz, forbidding fast adiabatic two-qubit gates.

One feature worth noting from Fig.~\ref{fig:FIG 3}(b) is the difference between the qubit-qubit detuning and the qubit anharmonicity, i.e.\ $\Delta_{12}-|\alpha_{\rm q}|$.
The bigger their difference, the flatter (smaller curvature) the parabola. Since flatness indicates the sensitivity to variations in $\widetilde{g}$, this means that, by designing $\Delta_{12}$ to be either considerably greater than $|\alpha_{\rm q}|$ or close to zero, $|\zeta^{\rm idle}|$ can be made robust against variations in circuit parameters such as qubit and coupler frequencies, the fabrication outcome of which usually differ from their design values by a few hundred megahertz. Even if there are sophisticated methods to improve accuracy~\cite{zhang2020high,hertzberg2020laser}, the performance at large scale is still under question.

To summarize, $\widetilde{g}=0$ is a universal condition for optimal residual ZZ suppressioin in the tunable-coupling architecture. Residual ZZ interaction $|\zeta^{\rm idle}|$ can be made sufficiently small if not zero, and robust against fabrication variations when the qubit-qubit detuning differs considerably from the qubit anharomonicity. However, as we will discuss in Sec.~\Rmnum{3}B, the large-detuning regime is preferred for attaining high-fidelity controlled-phase gates.

\section{Gate Performance Optimization}

In Sec.~\Rmnum{2}B, we explained the mechanism of coupler-assisted ZZ interaction. By adiabatically adjusting the coupler frequency from an idling (low-$|\zeta|$) bias into the near-resonance regime and back, a nontrivial phase accumulates on the $|101\rangle$ state, fulfilling a controlled-phase gate,
\begin{equation}
U = \begin{pmatrix} &1 & 0 & 0 & 0 \\
& 0  & e^{-i \phi_{2}} & 0 & 0 \\
&0 & 0 & e^{-i \phi_{1}} & 0 \\
&0 & 0 & 0 & e^{-i (\phi_{1}+ \phi_{2}+ \phi_{\rm zz})} \\
\end{pmatrix},
\label{eq:unitary_operator}
\end{equation}
where $\phi_{i} = \int_{0}^{T_{\rm g}} \widetilde{\omega}_i (t) \mathrm{d} t$ ($i=1,2$) is the trivial single-qubit phase and $\phi_{\rm zz} =\int_{0}^{T_{\rm g}} \zeta(t) \mathrm{d} t$ is the nontrivial entangling phase integrated over the gate time $T_{\rm g}$. The time dependence of $\zeta(t)$ is determined by the time dependence of the coupler frequency $\omega_{\rm c}(t)$ - basically the pulse shape - and the relation $\zeta(\omega_{\rm c})$ as shown in Fig.~\ref{fig:FIG 2}(b).
A controlled-Z (CZ) gate is obtained when $\phi_{\rm zz} = \pi$, after correcting for local single-qubit phases~\cite{dicarlo2009demonstration}.

Nonadiabatic (Landau-Zener) transitions during a finite-duration pulse are inevitable.
Among them, leakage transitions to non-computational states are particularly important because such errors - causing erroneous subsequent gate operations - are a type of correlated error which cannot be corrected by error-correction codes like surface codes~\cite{fowler2012surface}.
For better adiabaticity, a slower pulse is desirable. However, one also prefers a faster pulse for fewer decoherence errors.
A trade-off has to be made to maximize the overall performance, the gate fidelity.
In this section, we discuss how to implement fast and robust adiabatic C-phase or CZ gates through optimizing pulse shape and circuit design.
Here we pay attention to robustness, because, for scalable implementation, robustness is often more important than the highest achievable performance.
In this work, we do not touch upon nonadiabatic or diabatic gate schemes which heavily rely on control precision~\cite{li2019realisation,barends2019diabatic,negirneac2020high}.

\subsection{Adiabatic pulse engineering}

\begin{figure*}
	\includegraphics[width=18cm]{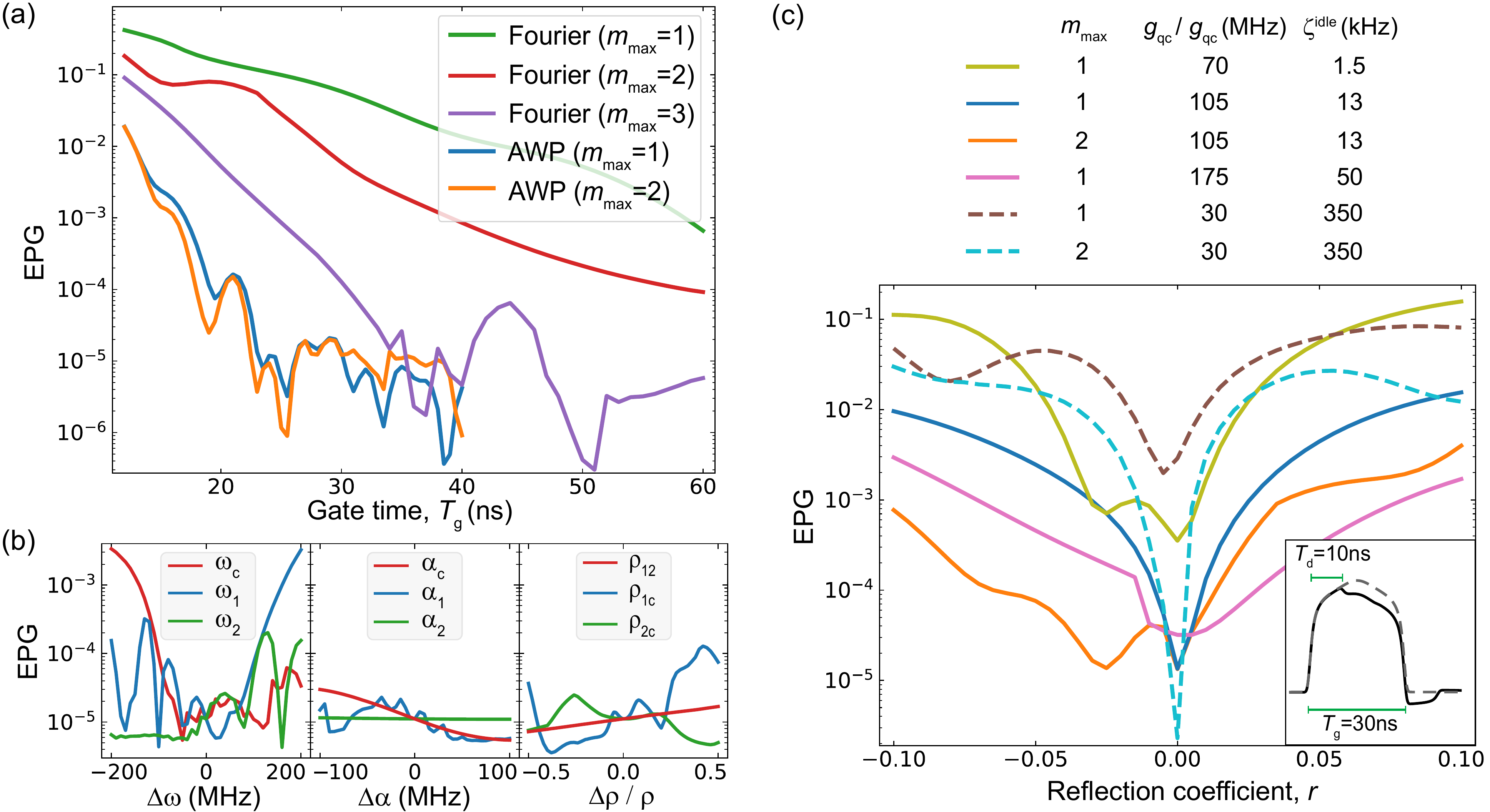}
	\caption{
		Simulated nonadiabatic error of CZ gates. Unless specified, the circuit parameters shared in all figures are: $\omega_{1,2} / 2\pi=$6.0, 5.4~GHz, $\alpha_{1,2,\rm c} / 2\pi=$-250, -250, -300~MHz, $\rho_{1\rm c,2\rm c,12}=0.018, 0.018, 0.0015$.
		(a) EPG versus gate time for different pulse shapes including Fourier basis and AWP with different component numbers.
		The coupler idles at $\omega_{\rm c} / 2\pi = 7.87$~GHz for minimizing $|\zeta|$.
        For each gate time, we numerically search for the minimum error rate using the Nelder-Mead method.
		(b) EPG of 30-ns CZ gate generated from one-component AWP versus deviation in frequencies (left), anharmonicities (center), and coupling coefficients (right) respectively with respect to the configuration in (a).
		(c) EPG of 30-ns CZ gate with distorted waveform. 
		As for distortion, we consider a reflected pulse added onto the original waveform with a delay time $T_{\rm d}=10$~ns and a varying reflection coefficient $r$.
		Inset is an example pulse waveform before (dashed curve) and after (solid curve) distortion. 
		We include a few configurations targeting different qubit-coupler coupling strength, $g_{1\rm c,2\rm c} / 2\pi \approx g_{\rm qc} =$70~MHz ($\rho_{\rm 1c,\rm 2c,12} = 0.012, 0.012, 0.0006$; solid light green line), 105~MHz ($\rho_{\rm 1c,\rm 2c,12} = 0.018, 0.018, 0.0015$; solid blue and orange line), and 175~MHz ($\rho_{\rm 1c,\rm 2c,12} = 0.03, 0.03, 0.0036$; solid pink line).
        The resulting residual ZZ interaction $\zeta^{\rm idle}$ at $\omega_{\rm c} / 2\pi \approx 8$~GHz is also listed for each case.
        For comparison, we also plot the results from the traditional coupler-free design (dashed lines).
        We assume $\omega_{1,2} / 2\pi = 6.0, 8.0$~GHz when idling ($\omega_{2}$ is tunable) and the qubit-qubit coupling strength $g_{\rm qq} / 2\pi=30$~MHz ($\rho_{12} = 0.005$).
        Considering the finite bandwidth of control electronics, all pulses are filtered by a Gaussian low-pass filter with a 300-MHz cut-off frequency.
		\label{fig:FIG 4}}
\end{figure*}

In this work, we propose a generally applicable and practically convenient method for finding the adiabatic pulse shape.
The method is inspired by the conventional fast adiabatic pulse used in the case of two-level system~\cite{martinis2014fast}), and generalizes it to the realm of multilevel.
For a given gate time $T_{\rm g}$, the time-derivative of a pulse $\omegac(t)$ can be decomposed into a number of $m_{\rm max}$ Fourier components,
\begin{equation}
\frac{\mathrm{d}\omega_{\rm c}}{\mathrm{d} t} = \frac{1}{\dw }\sum_{m=1}^{m_{\rm max}} \lambda_{m} \mathrm{sin}(\frac{2\pi mt}{T_{\rm g}} ), \label{eq:awp}
\end{equation}
with an additional prefactor $1/\dw$.
The definition of $\dw$, namely the $D$-factor, is
\begin{equation}
\dw = \sum_{{\rm s} \in \mathcal{S}} \sum_{{\rm s'} \neq {\rm s}}  \left| \frac{\langle {\rm s'} | \dot{{\rm s}} \rangle}{\omega_{{\rm s}}-\omega_{{\rm s'}}} \right|, \label{eq:ad_factor}
\end{equation}
where $|\dot{{\rm s}}\rangle = {\rm d} |{\rm s}\rangle /\rm d \omega_{\rm c}$. $\dw$ is a general measure of how diabatic an instance is. Greater $\dw$ means worse adiabaticity. It only depends on prior knowledge about the energy levels, and can be calculated with results from preliminary characterization measurements.
The pulse shape or waveform $\omega_{\rm c}(t)$ is obtained by iteratively solving the differential equation.
The idea behind Eq.~(\ref{eq:awp}) is simple: to dynamically adjust the slope of the pulse according to the instantaneous $D$-factor inversed in search for a relatively flatten and smooth response over the whole adiabatic process.
We thus name it adiabatically weighted pulse (AWP).

Using the AWP method, we optimize pulse parameters $\lambda_m$ for a CZ gate according to error rate per gate (EPG) defined by
\begin{equation}
\epsilon_{\rm CZ} = 1- \left| \frac{1}{4} \mathrm{Tr} ( U^{\dagger}_{\rm CZ} U_{\rm g}) \right|^2,
\label{eq:error_definition}
\end{equation}
where $U_{\rm CZ}$ is the target CZ gate unitary and $U_{\rm g}$ is the simulated unitary.
The simulation is done without decoherence, so the resulting gate errors are mostly nonadiabatic errors.
EPG for various gate times and different orders of expansion are plotted in Fig.~\ref{fig:FIG 4}(a), and are compared with those derived by simple Fourier expansion ($\dw=1$).
It is shown that nonadiabatic errors about $10^{-5}$ can be consistently achieved with gate time above 24~ns by using AWP with only one component, outperforming the simple Fourier pulse with multiple components. 
We also try adding extra components in AWP, but does not observe appreciable difference.
Therefore, it is convenient to use single-component AWP which has only one parameter to calibrate in experiment, drastically simplifying the calibration procedure.

Next, we study the robustness of the AWP method.
One cause of nonideal performance with AWP is the imprecise and incomplete knowledge about the system.
On one hand, experimental limitations may prevent us from obtaining the precise information about the system Hamiltonian. 
On the other hand, in a multi-qubit device, stray couplings to unwanted or environmental modes add to variations in observed quantities such as qubit or coupler frequencies.
Figure \ref{fig:FIG 4}(b) shows how gate error or adiabaticity is affected by deviation of certain circuit parameters.
We find that the AWP performance is insensitive to most parameters except to $\omega_{1}$ and $\omega_{\rm c}$.
In either case, one would underestimate the $D$-factor at the first avoided crossing which is the most nonadiabatic region during the adiabatic pulse.
After all, the gate error is kept at a very low level ($\sim\!10^{-5}$) given realistic parameter deviation: $\Delta \omega_i,\Delta \alpha_{i} \approx 10$ MHz and $\Delta \rho_{ij} /\rho_{ij} \approx 0.1$ ($i,j=1,2,{\rm c}$ and $i\neq j$).

Another common cause of lower-than-expected fidelity is pulse distortion, and one most important type of distortion is reflection.
For example, mismatch at two nodes along the signal line may produce a delayed reflection pulse added onto the original waveform (Fig.~\ref{fig:FIG 4}(c) inset). Such unsmooth distortion is the most harmful to an adiabatic process and extremely hard to characterize because of its short-time-scale nature~\cite{gustavsson2013improving,foxen2018high,jerger2019situ,rol2020time}. 
Its influence is illustrated in the example of traditional coupler-free design (Fig.~\ref{fig:FIG 4}(c), dashed lines). Due to the relatively small gap at the avoided crossing, the adiabatic gate is strongly sensitive to such distortion.
In contrast, AWP is more robust, thanks to the much larger gap in the tunable-coupling design (Fig.~\ref{fig:FIG 4}(c), solid lines). Actually, increasing the gap - that is, the qubit-coupler coupling $g_{q{\rm c}}$ - effectively lowers the nonadiabatic errors. However, the cost is stronger residual ZZ interaction that scales even faster ($\propto g_{q{\rm c}}^4$).

Our AWP method addresses adiabaticity in a multilevel system, which is necessary for the tunable-coupling design where more than one transition has to be considered simultaneously~\cite{sung2020realization}.
Compared to methods based on brute-force search~\cite{nobauer2015smooth,poggiali2018optimal,machnes2018tunable,li2019realisation}, our method is more scalable due to its simplicity in calibration and due to its robustness against parameter variation and pulse distortion.

\subsection{Circuit design optimization}

\begin{figure*}
    \includegraphics[width=18cm]{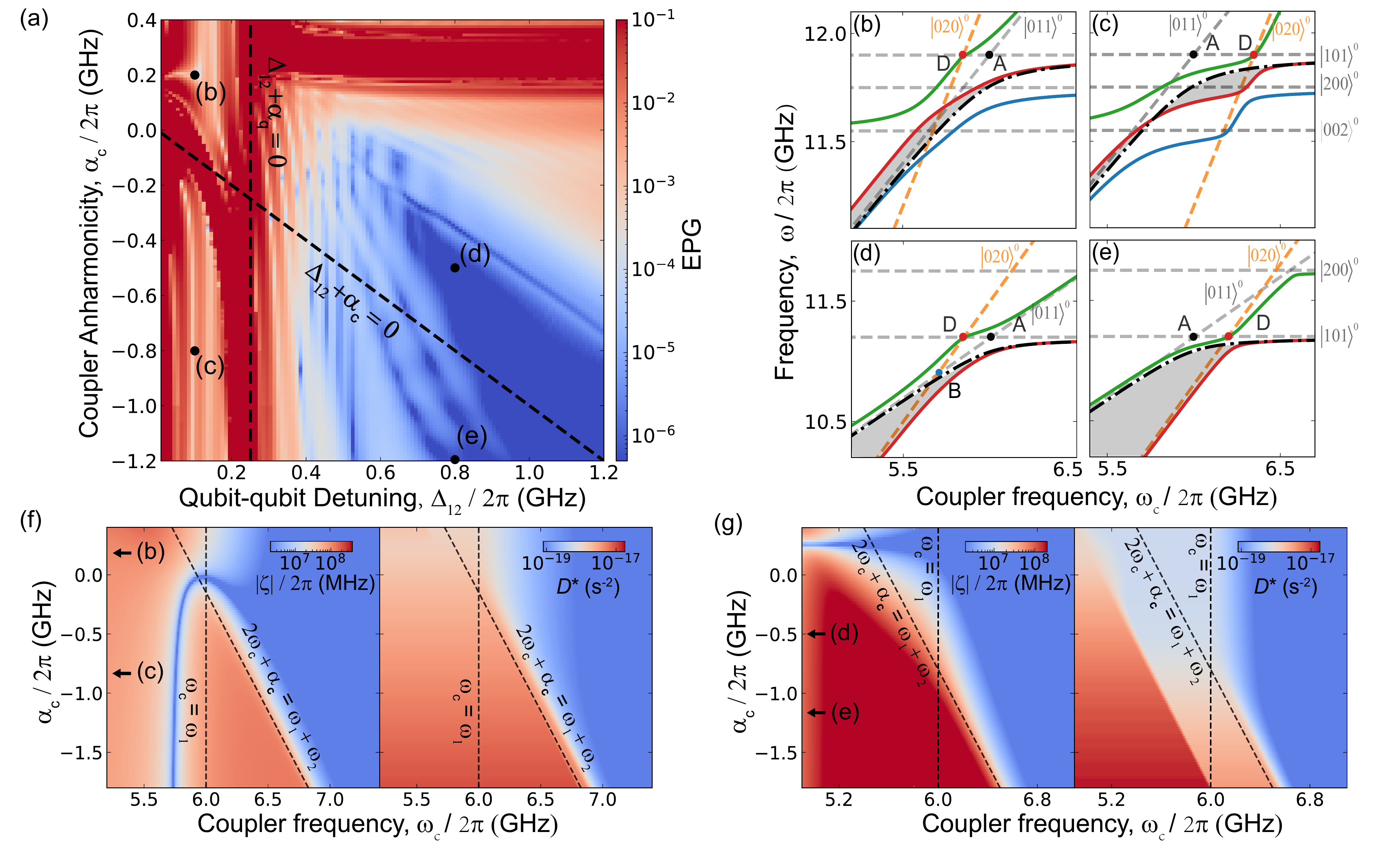}
    \caption{ 	
        (a) Optimized CZ gate error using one-component AWP ($T_{\rm g}=30$~ns) versus the qubit-qubit detuning $\Delta_{12}$ and coupler anharmonicity $\alpha_{\rm c}$. The circuit parameters are: $\omega_{1} / 2\pi = 6.0$~GHz, $\alpha_{1,2}/2\pi = -250$~MHz and $\rho_{\rm 1c,\rm 2c,\rm 12} = 0.018, 0.018, 0.0015$. Two dashed lines indicate the condition of $\Delta_{12}+\alpha_{\rm c}=0$ and $\Delta_{12}+\alpha_1=0$ respectively, dividing the plot into four different regions.
        (b-e) Level diagrams of the four typical cases marked in (a). The circuit parameters are $\Delta_{12}/2\pi = 100$~MHz, $\alpha_{\rm c}/2\pi = 200$~MHz for the case in (b); $\Delta_{12}/2\pi = 100$~MHz, $\alpha_{\rm c}/2\pi = -800$~MHz for the case in (c); $\Delta_{12}/2\pi = 800$~MHz, $\alpha_{\rm c}/2\pi = -500$~MHz for the case in (d); $\Delta_{12}/2\pi = 800$~MHz, $\alpha_{\rm c}/2\pi = -1200$~MHz for the case in (e). Point A and B indicate the same level crossings as in Fig.~\ref{fig:FIG 2}(a). Point D indicates the crossing between $|101\rangle^0$ and $|020\rangle^0$.
        (f-g) Magnitude of ZZ interaction $|\zeta|$ (left panel) and the \textit{maximum} instantanous $D$-factor $D^*(\omega_{\rm c})$ (right panel) versus the coupler frequency $\omega_{\rm c}$ and anharmonicity $\alpha_{\rm c}$ in the small-detuning regime ($\Delta_{12}/2\pi = 100$ MHz) and in the large-detuning regime ($\Delta_{12}/2\pi = 800$ MHz).
        The two dashed lines indicate the conditions of crossing D and A respectively.
        The arrows indicate the coupler anharmonicities used in (b-e).
    \label{fig:FIG 5}}
\end{figure*} 

In this section, we discuss how to exploit our gate scheme through optimizing circuit parameters.
In Eq.~(\ref{eq:hamiltonian_full}), there are 7 non-tunable parameters in a given circuit: $\Delta_{12},\alpha_1,\alpha_{\rm c},\alpha_2,\rho_{\rm 1c},\rho_{\rm 2c},\rho_{12}$ (the absolute value of $\omega_1$ or $\omega_2$ is not important here).  
Among them, we have the following observations. 
(\rmnum{1}) The coupling coefficients $\rho_{ij}$ are determined from the circuit layout. Although stronger coupling can improve gate performance, it makes residual ZZ interaction even worse. 
Besides, too strong a coupling may also suggest strong stray couplings to neighboring qubits.
We thus think that $g_{\rm qc}\approx100$~MHz is an appropriate choice at present.
(\rmnum{2}) The qubit anharmonicities ($\alpha_{1}$ and $\alpha_{2}$) typically range from $-300$~MHz to $-200$~MHz, balancing concerns about coherence and leakage.
(\rmnum{3}) The qubit-qubit detuning $\Delta_{12}$ and the coupler anharmonicity $\alpha_{\rm c}$ are considered free parameters to optimize. Both are deciding factors for the spectral structure in the near-resonance regime. They are expected to have strong influence on the gate performance.

Simulated EPG versus $\Delta_{12}$ and $\alpha_{\rm c}$ are shown in Fig.~\ref{fig:FIG 5}(a).
It can be seen that nonadiabatic errors are better suppressed when the qubit-qubit detuning is large ($\Delta_{12}>600$~MHz), and is around the absolute value of the coupler anharmonicity ($\Delta_{12} + \alpha_{\rm c} \gtrsim 0$).

We now explain these observations. First, it is difficult to achieve high fidelity in the small-detuning regime ($\Delta_{12} < |\alpha_{\rm 1}|$), because, when the qubit frequencies are close to each other, no configuration allows the $|101\rangle$ state to follow the adiabatic path without interruption as described in Sec.~\Rmnum{2}B.
The $|101\rangle$ state will either collide with $|200\rangle^0$ right after passing by $|011\rangle^0$ (Fig.~\ref{fig:FIG 5}(b)) or even collide with $|020\rangle^0$ before $|011\rangle^0$ (Fig.~\ref{fig:FIG 5}(c)). In both cases, the adiabatic path is interrupted by an weakly interacting state, leading to strong nonadiabatic errors.

In the large-detuning regime, if $\Delta_{12} + \alpha_{\rm c} > 0$, crossing D (formed by $|101\rangle^0$ and $|020\rangle^0$) is on the left side of crossing A (Fig.~\ref{fig:FIG 5}(d)). This configuration leads to the same ideal adiabatic path as in Fig.~\ref{fig:FIG 2}(a).
Otherwise, crossing D is on the right side of A (Fig.~\ref{fig:FIG 5}(e)), which means that the $|101\rangle$ state will first collide with $|020\rangle^0$, which is a small-gap crossing.
This explains why the regime of $\Delta_{12} + \alpha_{\rm c} \gtrsim 0$ is generally favored, as shown in Fig.~\ref{fig:FIG 5}(a).

The search for optimal circuit parameters may be guided by proper indicators without having to perform time-domain simulation with a presumed pulse shape.
In our scheme, two most relevant indicators are the achievable ZZ interaction $|\zeta(\omegac)|$ and the $D$-factor $\dw$. The former tells how fast the gate can be, relating to decoherence errors. The latter is a measure of the instantaneous adiabaticity, relating to nonadiabaticity error rate. Since nonadiabatic transitions are nonlinear effects, the total nonadiabatic error is usually determined by the worst instance during a gate. Therefore, we think that $D^*(\omegac) = \max\limits_{\omegac \leq \omegac' \leq \omegac^{\rm idle}} D(\omegac')$, the \textit{maximum} instantaneous $D$-factor when adjusting the coupler frequency from the idling point to $\omegac$, is a more convenient indicator.

Combining $|\zeta|$ and $D^*$, it becomes easier to identify a good configuration in the parameter space.
In Fig.~\ref{fig:FIG 5}(f-g), we plot the two indicators versus $\alpha_{\rm c}$ and $\omegac$ for both small-detuning and large-detuning case.
Ideally, one would prefer a set of configuration that gives large $|\zeta|$ (as red as possible) and small $D^*$ (as blue as possible).
Obviously, there is no acceptable choice in the small-detuning case. Regions with large $|\zeta|$ (red) also have large $D^*$ (red).
In the large-detuning case, large $|\zeta|$ ($\sim100$~MHz) can be achieved in the range of $-800~\textrm{MHz} \leq \omegac \leq -400~\textrm{MHz}$ while $D^*$ is maintained at a relatively low level ($<10^{-18}~s^{-2}$).
This is consistent with what is shown in Fig.~\ref{fig:FIG 5}(a).

To summarize, large qubit-qubit detuning is favorable for better gate fidelity and small, robust residual ZZ interaction as discussed in Sec.\Rmnum{2}.C. The magnitude of the coupler anharmonicity should be equally large or moderately smaller.

\subsection{Gate error caused by stray couplings}

\begin{figure}
	\includegraphics[width=8.5cm]{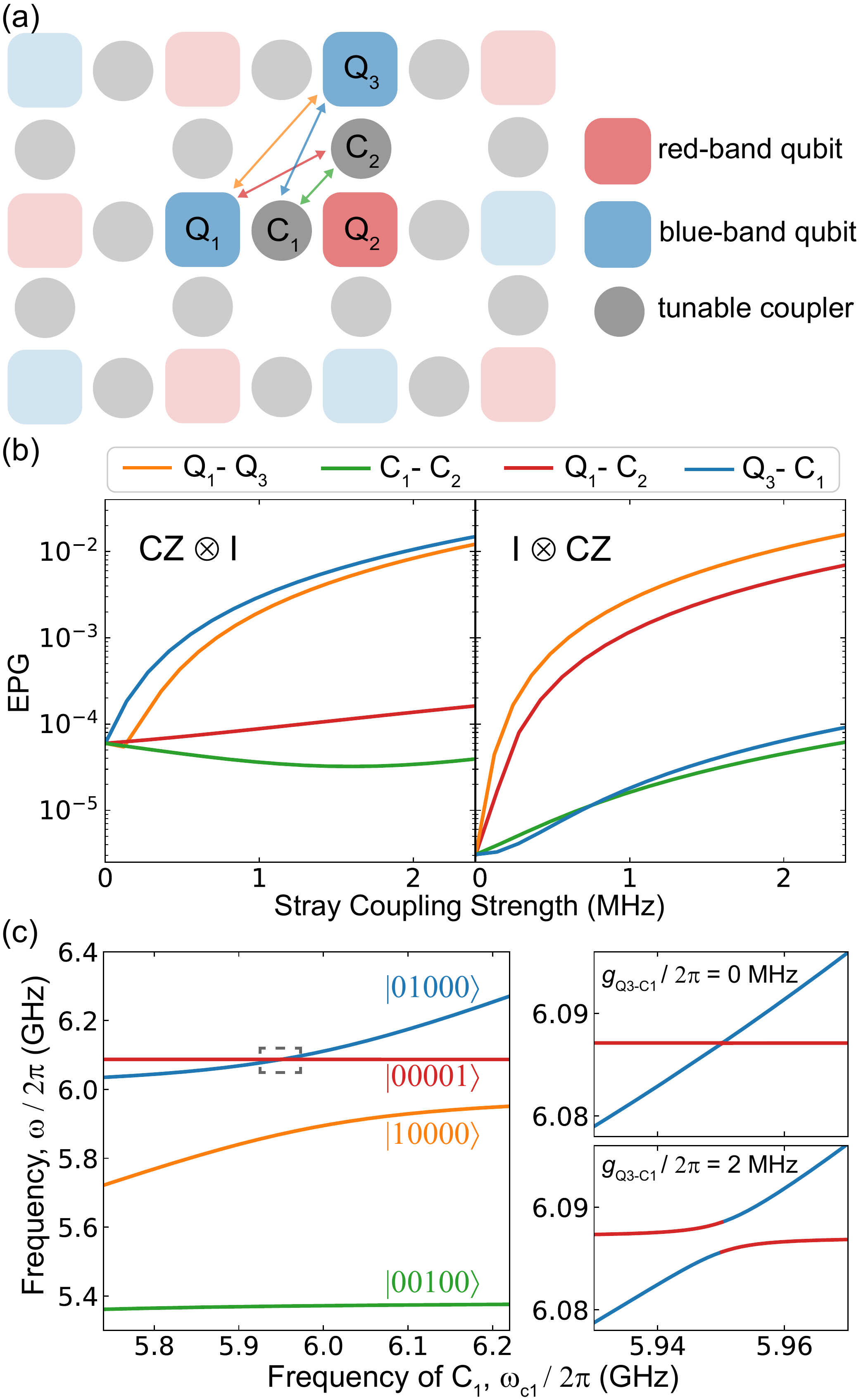}
	\caption{ 	
		(a) Layout of 2D qubit array in a square lattice. Red and blue squares representing qubits in different frequency bands are alternately arranged to keep large detuning between neighboring qubits. Gray circles denote tunable couplers. Arrows indicate stray couplings. 
		(b) Error rates of $\rm CZ \otimes I$ (left) and $\rm I \otimes CZ$ (right) versus stray coupling strength. 
		Different colors correspond to different types of stray coupling as listed on top. 
		The parameters for the five-mode circuit are: $\omega_{1,2,3}/2\pi= 6.0, 5.4, 6.1$ GHz, $\alpha_{\rm q,\rm c}/2\pi = -250, -300$ MHz and $\rho_{\rm qc, qq} = 0.018, 0.0015$. 
		Gate time is 30~ns.
		The AWP method is used in pulse finding.
        (c) Spectra of energy levels in the single-excitation manifold (left panel). Zoom-in of the dashed area is shown for the case of zero (top right) and finite (bottom right) stray coupling between $\rm Q_3$ and $\rm C_1$. 
		\label{fig:FIG 6}}
\end{figure} 

To end this section, we emphasize the effect from stray coupling.
Consider a 2D qubit array for implementing surface code, as shown in Fig.~\ref{fig:FIG 6}(a). As discussed above, we generally prefer the large-detuning regime. This allows us to alternate the red-band (lower-frequency) and blue-band (higher-frequency) qubits.
Such an arrangement provides intrinsic robustness against nearest-neighbor crosstalk for both XY-line and Z-line signals~\cite{xu2020high}. 
However, longer-distance stray couplings may nevertheless cause severe phase error and leakage~\cite{zajac2021spectators}.

Here, we consider all types of stray coupling (qubit-qubit, coupler-coupler, qubit-coupler) in the subcircuit highlighted in Fig.~\ref{fig:FIG 6}(a).
We simulate error rates of the CZ gate between $\rm Q_1$ and $\rm Q_2$ ($\rm CZ \otimes I$), as well as between $\rm Q_2$ and $\rm Q_3$ ($\rm I \otimes CZ $).
The dependence of EPG on the coupling strength for each type of stray coupling is plotted in Fig.~\ref{fig:FIG 6}(b).
The results show that the next-nearest-neighbor qubit-qubit and qubit-coupler stray couplings have stronger impact on gate errors than the coupler-coupler coupling. 
To explain it, we zoom in the spectra of the single-excitation manifold in this five-mode system ($|n_{\rm q1} n_{\rm c1} n_{\rm q2} n_{\rm c2} n_{\rm q3}\rangle$), as shown in Fig.~\ref{fig:FIG 6}(c). 
When performing a CZ gate between $\rm Q_1$ and $\rm Q_2$, the computational state $|00001\rangle$ (the excited state of $\rm Q_3$) will collide with the non-compuational state $|01000\rangle$ (the excited state of $\rm C_1$).
In the case without stray coupling, the crossing has a negligible gap (<100~kHz). Due to the relatively short timescale of the gate pulse (usually a few tens of nanoseconds), to sweep through this crossing is nearly a fully diabatic process, keeping the system at the wanted state ($|00001\rangle$).
On the other hand, having a stray coupling of a few megahertz is the most harmful to this process, as it stays halfway between the diabatic and adiabatic limit.

\section{Noise and Decoherence}

\begin{figure*}
	\includegraphics[width=18cm]{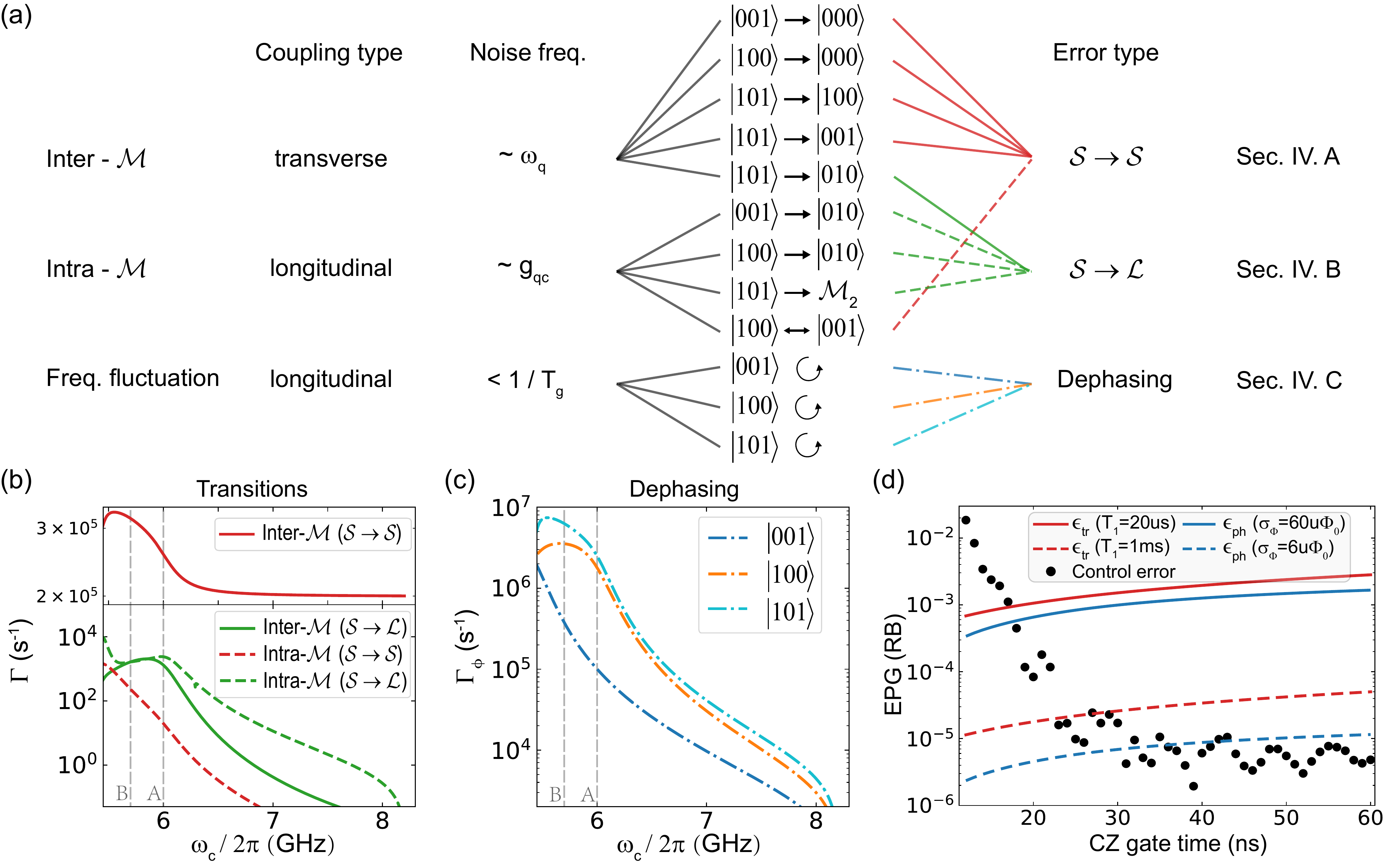}
	\caption{ 	
		(a) Categorization of major decoherence processes according various properties.
        The coupling axis refers to the direction along which the noise couples to the corresponding mode (isolated).
        Noise frequency indicates the relevant spectral components that contribute to the corresponding decoherence processes.
        Most inter-manifold transitions are transitions within the computational subspace ($\mathcal{S} \rightarrow \mathcal{S}$) with the exception of $|101\rangle \to |010\rangle$.
        Most intra-manifold transitions are leakage transitions ($\mathcal{S} \rightarrow \mathcal{L}$) with the exception of $|100\rangle \leftrightarrow |001\rangle$.
		(b) Calculated instantaneous transition rates versus the coupler frequency for various types of transitions.
        The coupler is assumed to be a split transmon with a maximum frequency of 8.2~GHz. Other parameters are the same as the ones used in Fig.~\ref{fig:FIG 4}(a).
        Both qubits have $T_1=20~\mu$s. The coupler $T_1$ is assumed to be shorter, $10~\mu$s.
        For longitudinal noise, we assume flux noise with a $1/f$-type spectrum, $S_{\Phi}(\omega) = A_{\Phi} / (|\omega|/2\pi)$, with $A_{\Phi} = (10~\mu\Phi_0)^2$.
        Vertical lines are references to level crossing A and B, same as those introduced in Fig.~\ref{fig:FIG 2}(a). 
        (c) Calculated instantaneous pure-dephasing rates versus the coupler frequency.
		We assume a total quasistatic flux fluctuation of $\sigma_{\Phi} = 60~\mu\Phi_0$ which corresponds to the same $1/f$ noise used above.
		(d) CZ gate errors in RB metric versus gate time for different decohering processes.
    	Noise-induced errors - transitional (solid red line) and phase (solid blue line) - are calculated from the model described in Eq.~(\ref{eq:error_rb}-\ref{eq:phase_error}) using the same noise assumed above.
		Also shown is the case assuming $T_1=1$~ms (dashed red line) and $\sigma_{\Phi} = 6~\mu\Phi_0$ (dashed blue line). 
        Coherent (nonadiabatic) errors are from numerically simulation.
		\label{fig:FIG 7}}
\end{figure*} 

In this section, we discuss relevant decoherence phenomena associated with our gate scheme.
In Fig.~\ref{fig:FIG 7}(a), we categorize major decohering processes into three types: inter-manifold transitions ($\mathcal{M}_i \to \mathcal{M}_j, i \neq j$), intra-manifold transitions ($\mathcal{M}_i \to \mathcal{M}_i$), and frequency fluctuations (no transitions).
The starting state is always one of the computational basis states in $\mathcal{S}$.
In the following sections, we will study each type and summarize them in a combined error model connecting to randomized benchmarking experiments.

\subsection{Inter-manifold transitions}

In this system, most transitions between neighboring energy bands (excitation number differed by 1) are induced by the noise transversally coupled to qubits or couplers, the same noise that leads to relaxation and excitation.
The instantaneous transition rate from state $|{\rm s}\rangle=|{\rm s}(\omegac)\rangle$ to $|{\rm s'}\rangle=|{\rm s'}(\omegac)\rangle$ follows
\begin{equation}
    \begin{aligned}
    \Gamma(\omegac) &=  \frac{1}{2}   \sum_{i=1,2,\rm c} \left|\langle {\rm s'}| a_i+a_i^{\dagger} |{\rm s}\rangle  \right|^2 S_{i,\perp}(\omega_{ {\rm s} }-\omega_{{\rm s'}}), \label{eq:rate_inter}
    \end{aligned} 
\end{equation}
where $S_{i,\perp}(\omega)$ is the power spectral density of noise that transversally couples to mode $i$.

Here, we consider only relaxation processes, as excitation is exponentially suppressed due to the low-temperature environment, typically a few tens of milliKelvin.
Assuming frequency-independent $T_1$ (white noise spectrum) for each mode, we calculate the instantaneous or $\omegac$-dependent transition rates as shown in Fig.~\ref{fig:FIG 7}(b). We group them into two error types, within the computational subspace ($\mathcal{S} \to \mathcal{S}$) and leakage ($\mathcal{S} \to \mathcal{L}$). In this case, there is only one leakage transition, $|101\rangle \to |010\rangle$.
It can be seen that, for the inter-manifold case, the transition rates within the computational subspace $\Gamma_{\mathcal{S} \to \mathcal{S}}$ (solid red line) are more than two orders of magnitude higher than the leakage rates $\Gamma_{\mathcal{S} \to \mathcal{L}}$ (solid green line), because the matrix elements for $\Gamma_{\mathcal{S} \to \mathcal{S}}$ transitions in Eq.~(\ref{eq:rate_inter}) are on the order of unity, while the transition between $|101\rangle$ and $|010\rangle$ is indirect, leading to small matrix elements.
The $\omegac$-dependencies reflect how diabatic states participate in the instantaneous energy eigenstates. In the shown example, we deliberately assume a lower $T_1$ for the tunable coupler, so that ${\Gamma}_{\mathcal{S} \to \mathcal{S}}$ becomes higher in the near-resonance regime, suggesting more contribution from the coupler relaxation.

\subsection{Intra-manifold transitions}

Transitions within the same energy band may be caused by longitudinal noise, which, in this system, is mostly the flux noise in the coupler.
The instantaneous transition rate follows
\begin{equation}
    \begin{aligned}
    \Gamma(\omegac) &= \frac{1}{2}  \left|\langle {\rm s'}| 2\ac^{\dagger} \ac | {\rm s} \rangle  \right|^2 S_{\rm c,\parallel}(\omega_{ {\rm s} }-\omega_{ {\rm s'}}), \label{eq:rate_intra}
    \end{aligned} 
\end{equation}
where $S_{\rm c,\parallel}(\omega)$ is the power spectral density of noise that longitudinally couples to the tunable coupler.

Assuming that the flux noise has a $1/f$-like PSD which is $(10~\mu\Phi_0)^2$ at 1~Hz ($\Phi_0$ is the superconducting flux quantum), we calculate the transition rates and show them in Fig.~\ref{fig:FIG 7}(b).
Similar to the inter-manifold case, we group them into the $\mathcal{S} \to \mathcal{S}$ type - there is only a (two-way) transition, $|100\rangle \leftrightarrow |001\rangle$ (dashed red line) -  and the $\mathcal{S} \to \mathcal{L}$ type (dashed green line).
These rates become significant only when the initial and final states hybridize at their avoided crossings with increased matrix elements in Eq.~(\ref{eq:rate_intra}) and smaller frequency gap (hence larger noise PSD because of the $1/f$ dependence).
Apparently, all these transition rates are considerably lower than those from the $T_1$ effect (solid red line), so this type of transition is not yet the limiting factor. 

Although the noise-induced leakage rates seem to be low, it is of particular importance because state leakage can be a source of correlated errors that make error correction codes fail, hindering fault-tolerant quantum computation~\cite{ghosh2013understanding,terhal2015quantum,kelly2015state,mcewen2021removing}.

\subsection{Pure dephasing}

The longitudinal noise also induces pure dephasing.
Assuming quasistatic flux noise $\delta\Phi$ (Gaussian distributed), the instantaneous self-dephasing rate for state $|{\rm s}\rangle$ is
\begin{equation}
    \Gamma_{\phi}^{\rm s}(\omegac) = \frac{\partial \widetilde{\omega}_{\rm s}}{\partial \omegac} \frac{\partial \omegac}{\partial \Phi} \, \sigma_\Phi/\sqrt{2}, \label{eq:rate_dephasing}
\end{equation}
where $\sigma_\Phi = \sqrt{\langle \delta\Phi^2 \rangle}$ is the standard deviation of flux fluctuation.
As shown in Fig.~\ref{fig:FIG 7}(c), the dephasing rates for $|100\rangle$ and $|101\rangle$ become greater ($>10^6~{\rm s^{-1}}$) in the near-resonance regime where they have considerable overlap with excited states of the coupler.
The $|101\rangle$ case is worse because it has strong overlap with $|020\rangle^0$ in which the spectral slope $\frac{\partial \widetilde{\omega}_{\rm s}}{\partial \omegac}$ is nearly doubled.
Note that, here in this case here, the noise has a same origin (the coupler flux), so the noise-induced quasistatic phases on different states are fully correlated.
The often-said (relative) dephasing rate between two states is derived by taking the difference, e.g. $|\Gamma_{\phi}^{\rm s}-\Gamma_{\phi}^{\rm s'}|$.
Note that pure dephasing is a type of unitary error. Its contribution to gate errors is drastically reduced in a random circuit such as the RB experiment.

\subsection{Noise-induced gate error model}

The model of the noise-induced two-qubit gate error connecting to the RB result is (see detailed derivation in Appendix C)
\begin{equation}
\begin{aligned}
&\epsilon_{\rm noise} = \epsilon_{\rm tr} + \epsilon_{\rm ph} , \quad \rm{where} \\
&\epsilon_{\rm tr} = \frac{1}{5} \int_{0}^{T_{\rm g}}  \Gamma_{\mathcal{S} \to \mathcal{S}} (t) \, \ud t  + \frac{1}{4} \int_{0}^{T_{\rm g}}  \Gamma_{\mathcal{S} \to \mathcal{L}} (t) \, \ud t  , \quad \rm{and} \\
&\epsilon_{\rm ph} = \frac{1}{20} \big[ 3 \langle (\Delta \phi_{100})^2 \rangle
+3 \langle (\Delta \phi_{001})^2 \rangle 
+3 \langle (\Delta \phi_{101})^2 \rangle  \\
&-2 \langle \Delta \phi_{001} \Delta \phi_{100} \rangle 
-2 \langle \Delta \phi_{001} \Delta \phi_{101} \rangle
-2 \langle \Delta \phi_{100} \Delta \phi_{101} \rangle \big], \label{eq:error_rb}
\end{aligned}
\end{equation}
where $\Gamma_{\mathcal{S} \to \mathcal{S}}(t)$ is the total transitional error rate within computational subspace accumulated during a gate, and $\Gamma_{\mathcal{S} \to \mathcal{L}}(t)$ is the total leakage error rate. The time dependence can be obatined from the relation $\Gamma(\omegac)$ (Fig.~\ref{fig:FIG 7}(b)) and the pulse shape $\omegac(t)$.
$\langle \Delta \phi_{{\rm s}} \Delta \phi_{{\rm s'}} \rangle $ is the (co)variance of the erroneous phases ($\Delta {\phi}$) accumulated on state ${\rm s}$ and ${\rm s'}$ during a gate.

The dephasing error model in Eq.~(\ref{eq:error_rb}) applies to noise with both long and short correlation time. However, since the majority of dephasing comes from low-frequency flux fluctuations, here we consider only the quasistatic flux noise.
Hence, the phase errors accumulated on different states are co-varying unitary errors. The average phase error rate is then
\begin{equation}
\begin{aligned}
\epsilon_{\rm ph} &= \frac{1}{10} \big[  3(\epsilon_{\phi}^{100})^2
+ 3(\epsilon_{\phi}^{001})^2 + 3(\epsilon_{\phi}^{101})^2 \\ 
& - 2\epsilon_{\phi}^{100}\epsilon_{\phi}^{001} - 2 \epsilon_{\phi}^{100}\epsilon_{\phi}^{101} - 2 \epsilon_{\phi}^{001}\epsilon_{\phi}^{101} \big] , 
\label{eq:phase_error}
\end{aligned}
\end{equation}
where $\epsilon_{\phi}^{{\rm s}} = \int_{0}^{T_{\rm g}} \Gamma_{ \phi }^{{\rm s}} (t) \, \ud t $ is the phase error for state $|{\rm s} \rangle$ during a gate.
$\Gamma_{ \phi }^{{\rm s}} (t)$ can be obtained from the relation $\Gamma_{ \phi }^{{\rm s}} (\omegac)$ (Fig.~\ref{fig:FIG 7}(c)) and the pulse shape $\omegac(t)$.
Equation~(\ref{eq:phase_error}) tells that the gate error due to dephasing is smaller for noise from the same origin - leading to covarying (same-direction) phases - than for independent noise sources.

In Fig.~\ref{fig:FIG 7}(e), we show the CZ gate error from different contributions versus gate time.
The noise-induced errors are calculated from Eq.~(\ref{eq:error_rb}).
The coherent control errors (black dots) are obtained by averaging the simulated errors over a total of 60 different two-qubit states.
It can be seen that, given the $T_1$ times assumed above ($20~{\rm \mu s}$ for the qubits; $10~{\rm \mu s}$ for the coupler), the $T_1$ noise is the leading error source of errors (solid red line) when the gate time is above about 15~ns.
The gate error due to pure dephasing in the RB metric is slightly lower (solid blue line) due to its unitary nature (quadratic relation), in spite of strong pure-dephasing rates in the near-resonance regime ($>\!10^6~{\rm s^{-1}}$).
Both $T_1$ and pure-dephasing processes are significant contributors when the gate time is 30~ns, at which the non-adiabatic error is only about $10^{-5}$.

However, given the state-of-the-art results, $T_1$ times are getting close to the 1-ms mark~\cite{place2021new,wang2021transmon} and local $1/f$ flux noise can be about $(1~\mu\Phi_0)^2$ at 1~Hz~\cite{braumuller2020characterizing}, the calculated transitional error is lowered to be around $10^{-5}$ (dashed red line) for 30~ns gate time, and the dephasing error is even lower (dashed blue line).
Therefore, our scheme is promising to achieve a total EPG at the $10^{-5}$ level, two-orders-of-magnitude better than the state-of-the-art results~\cite{arute2019quantum,stehlik2021tunable}, while inhering simplicity and robustness.

\section{Generalized control schemes}

\begin{figure}
	\includegraphics[width=8.5cm]{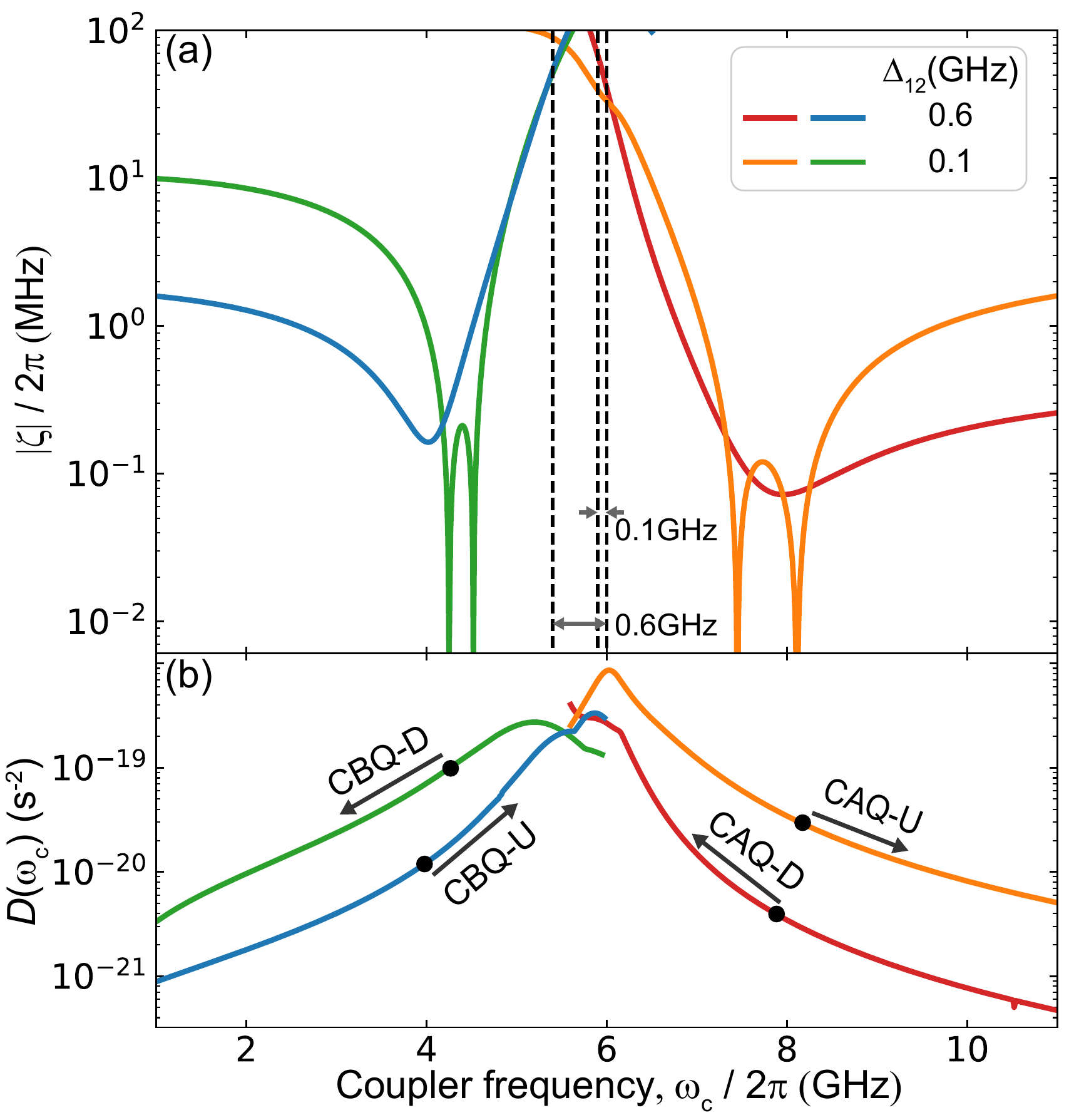}
	\caption{ 
		(a) Sketch of four coupler-assisted control schemes. 
		The solid lines represent absolute ZZ interaction strength.
		% with $\Delta_{12} = 0.6$ GHz (red and blue) and $\Delta_{12} = 0.1$ GHz (orange and green).
		Circuit parameters: $\omega_{1}/2\pi = 6.0$ GHz, $\omega_{2}/2\pi =5.4$ GHz for the large-detuning cases (red and blue), 5.9~GHz for the small-detuning cases (orange and green),
		$\rho_{\rm qc, qq} = 0.03, 0.004$ for the CAQ cases (red and orange), $\rho_{\rm qc, qq} = 0.04, -0.004$ for the CBQ cases (blue and green), and $\alpha_{1,2,\rm c}/2\pi = -250,-250,-300$ MHz. The valleys in the small detuning cases correspond to two roots of $\zeta = 0$, as discussed in Sec.\Rmnum{2}.C. (b) $\dw$ for the four cases. The black dots indicate where the coupler is idly biased.
		\label{fig:FIG 8}}
\end{figure}

In the previous sections, we have assumed that the coupling coefficients ($g_{\rm 1c}$, $g_{\rm 2c}$ and $g_{\rm 12}$) all have the same sign so that the coupler should idle above the qubit frequencies to suppress the residual coupling (Eq.~(\ref{eq:g_effective})). However, they may be engineered to have opposite signs as demonstrated in~\cite{stehlik2021tunable,sete2021floating}. Hence, one can idle the coupler below the qubit frequencies to implement this adiabatic CZ gate scheme by pulsing the coupler uptowards in an almost reversed process.
Also, we have assumed that the coupler is tuned towards the qubits into the near-resonance regime for strong ZZ interaction, but one can, in principle, tune the coupler away from the qubits to obtain significant ZZ interaction as well.
In this case, the coupler-mediated qubit-qubit coupling is suppressed, and the direct qubit-qubit coupling $g_{\rm 12}$ becomes dominant.
In this section, we extend our discussion to these generalized situations.
Accordingly, there are four cases: (\rmnum{1}) coupler-above-qubit and uptuning (CAQ-U); (\rmnum{2}) coupler-above-qubit and downtuning (CAQ-D); (\rmnum{3}) coupler-below-qubit and uptuning (CBQ-U); (\rmnum{4}) coupler-below-qubit and downtuning (CBQ-D).
We summarize pros and cons of each in Table 1 and discuss over them as follows. The discussion is based on an all-transmon circuit.

\begin{table}[h!]
	\begin{center}
		\begin{tabular}{|c|c|c|c|c|}
			\hline
			& \textbf{CAQ-U} & \textbf{CAQ-D} & \textbf{CBQ-U}  & \textbf{CBQ-D}\\
            \hline
			\textbf{\makecell[c]{ Achievable \\ ZZ interaction}} & weak & strong & strong & weak \\
			\hline
			\textbf{\makecell[c]{ Preferred \\ qubit detuning}} & small & large & large & small \\
            \hline
			\textbf{\makecell[c]{ Adiabaticity \\ condition}} & Good & Bad & Bad & Good \\
			\hline
			\textbf{\makecell[c]{ Strong \\ idling bias}} & Yes & No & Yes & No \\
			\hline
			\textbf{\makecell[c]{ Coupler \\ excitation}} & Less & Less & Less & More \\
			\hline
		\end{tabular}
	\end{center}
	\caption{Concerns of generalized C-phase gate schemes in transmon circuits.}
\end{table}	

(\rmnum{1}) ZZ interaction.
As shown in Fig.~\ref{fig:FIG 8}(a), $|\zeta|$ can be made strong enough for completing a CZ gate in tens of nanoseconds by tuning the coupler both towards and away from the qubits.
The maximum achievable $|\zeta|$ is more than an order of magnitude stronger in the former schemes - CAQ-D and CBQ-U - due to the coupler-assisted strong ZZ interaction described in the previous sections.
However, when tuning away the coupler in the CAQ-U and CBQ-D schemes, only the direct qubit-qubit coupling ($g_{12}$) survives, giving finite $|\zeta|$. 

(\rmnum{2}) Qubit-qubit detuning.
The optimal qubit-qubit detuning depends on the control scheme. As shown in Fig.~\ref{fig:FIG 8}(a), large detuning is favorable in the CAQ-D and CBQ-U schemes, while small detuning is favorable in the CAQ-U and CBQ-D schemes.
However, large detuing is useful for ameliorating the inflence from crosstalk by alternating qubit frequencies. 

(\rmnum{3}) Adiabaticity.
The cost for tuning the coupler into the near-resonance regime in CAQ-D and CBQ-U is the worse adiabaticity given the rising $D$-factor (Fig.~\ref{fig:FIG 8}(b)).
In contrast, without level crossings, adiabaticity is better protested in CAQ-U and CBQ-D, but the price to pay is the slower gate speed.

(\rmnum{4}) Idling bias. In both uptuning cases, a strong DC current is needed to flux-bias the coupler down to its idling frequency. Such DC current may be detrimental in scalable applications because it may easily heat up the dilution refrigerator which provides a low-temperature, low-noise measurement environment.

(\rmnum{5}) Coupler excitation. Excitation of the coupler can be considered a leakage transition, which completely destroys subsequent operations and, unfortunately, cannot be corrected by surface codes. In our example, an unwanted excitation of the coupler will result in an average CZ gate error about $\sim0.7$. 
It may occur during state preparation and gate operations. A higher coupler frequency generally helps reduce thermal excitations.
Therefore, this makes the CBQ-D scheme relatively unfavorable as it requires biasing the coupler to very low frequencies (<2~GHz). 
However, it is possible to reset the coupler by selective measurement~\cite{riste2012initialization} or dissipation engineering techniques~\cite{valenzuela2006microwave,mariantoni2011implementing,geerlings2013demonstrating,mcewen2021removing,zhou2021rapid}.

In general, a trade-off has to be made based upon limitations of actual hardware and measurement setup when deciding on an appropriate scheme.
In the near future, the CAQ-D scheme (the main example in this work) is advantageous as it strikes a good balance between nonadiabatic and decoherence errors (30~ns gate time, $10^{-5}$ nonadiabatic error)by virtue of our pulse engineering method (AWP).
However, in the long run, as coherence times keep improving, we may have more coherence budget to strike a better balance by elongating the gate duration. In that case, the CBQ-D scheme may also be appealing.

\section{Summary}

To summarize, we have discussed in detail the underlying mechanism of the coupler-assisted adiabatic controlled-phase gate previously demonstrated in~\cite{collodo2020implementation,xu2020high}.
Strong ZZ interaction arises from non-trivial level repulsion between double-excitation states with the help of the tunable coupler. 
Here, the idea of using an intermediate state for initiating interactions previously inaccessible is an inspiring method for engineering advanced quantum operations.
During idling, residual ZZ interaction can be suppressed in the dispersive regime by properly choosing the design parameters.
For such a multilevel system, we propose a convenient method for shaping an adiabatic pulse based on prior knowledge about the system Hamiltonian. Pulses generated by this method - using only a single calibrating parameter - have better adiabaticity and robustness.
Also, by choosing proper values for the qubit-qubit detuning and the coupler anharmonicity, we can further improve the gate fidelity. For an all-transmon circuit, large qubit-qubit detuning and equally large coupler anharmonicity are favorable.
We study noise and decoherence phenomena in this scheme and find that energy relaxation and flux-noise-induced dephasing are the major sources of errors.
Given state-of-the-art $T_1$ times and low-frequency flux noise, EPG may be reduced to be near $10^{-5}$, showing the great potential of this scheme in future scalable applications of quantum information processing.
By comparing different extension schemes, the CAQ-D scheme seems so far the best choice, given the more mature transmon technology.

\section*{ACKNOWLEDGMENTS}

We thank Orkesh Nurbolat, Jiawei Qiu and Yang Yu for insightful discussion.
This work was supported by the Key-Area Research and Development Program of Guang-Dong
Province (Grant No. 2018B030326001), the National Natural Science Foundation of China (U1801661), the Guangdong Innovative and Entrepreneurial Research Team Program (2016ZT06D348), the Guangdong Provincial Key Laboratory (Grant No.2019B121203002), the Natural Science Foundation of Guangdong Province (2017B030308003),
and the Science, Technology and Innovation Commission of Shenzhen Municipality (JCYJ20170412152620376, KYTDPT20181011104202253), and the NSF of Beijing (Grants No. Z190012).

\onecolumngrid
\newpage

\setcounter{figure}{0}
\renewcommand{\thefigure}{S\arabic{figure}}	

\setcounter{equation}{0}
\renewcommand{\theequation}{A\arabic{equation}}	

\section*{APPENDIX A: ZZ interaction from non-trivial level repulsion}

\begin{figure}
	\includegraphics[width=8.5cm]{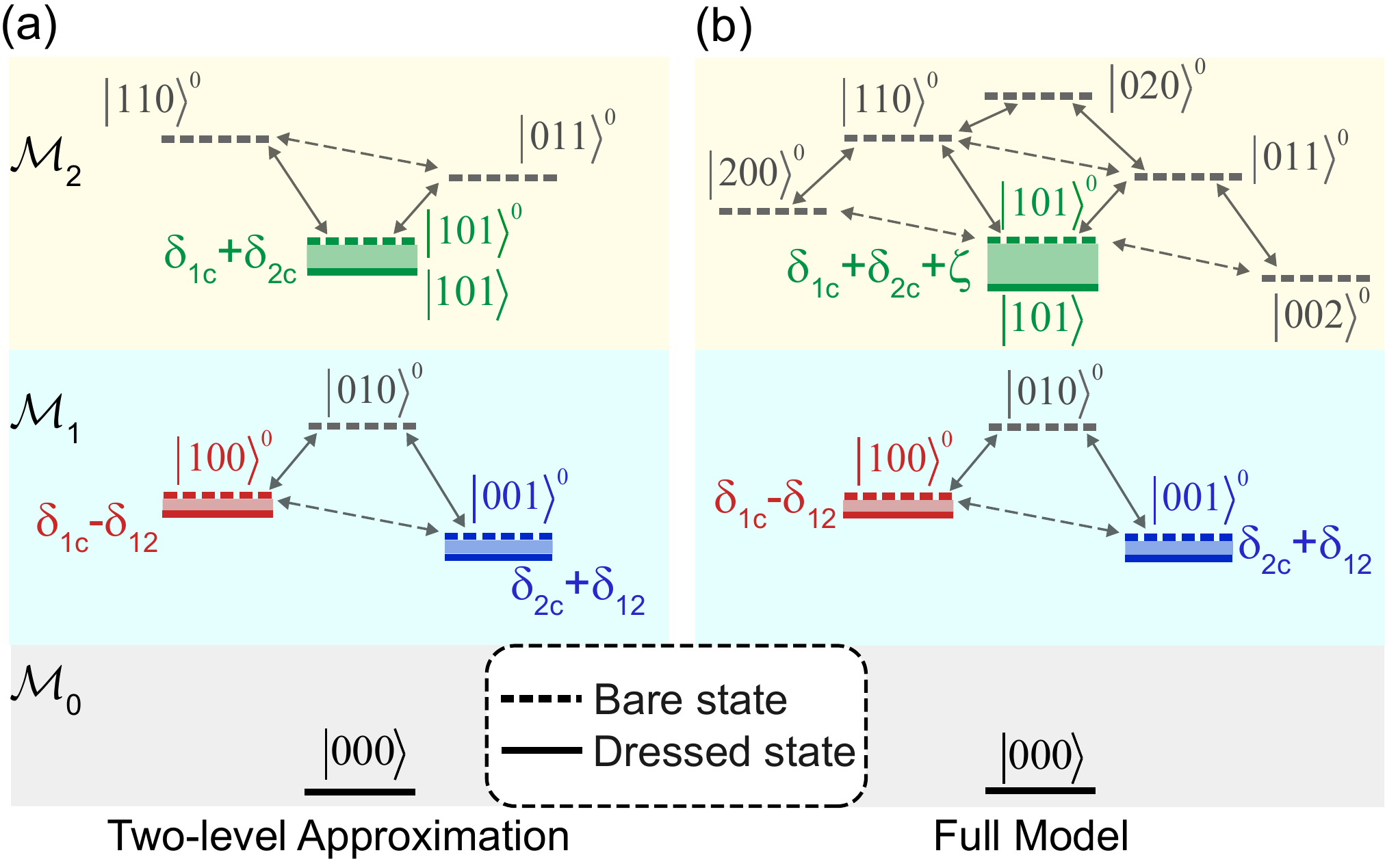}
	\caption{			
		Level structures of a tunable coupling system using two-level approximation (a) and full model (b). The lowest three manifolds are shown with gray ($\mathcal{M}_0$), cyan ($\mathcal{M}_1$) and yellow ($\mathcal{M}_2$) background.
		Solid and dashed arrows represent strong qubit-coupler coupling ($\propto g_{i \rm c},i=1,2 $) and weak qubit-qubit coupling ($\propto g_{12}$) respectively. The frequency shift of a computational state is illustrated as the rectangular between the corresponding adiabatic state (solid horizontal) and diabatic state (dashed horizontal).
		\label{fig:TLA}}
\end{figure}

Although there is no explicit ZZ interaction term in the Hamiltonian of the tunable coupling system (Eq.~(\ref{eq:hamiltonian_full}) in the main text), non-zero ZZ interaction still emerges due to the level repulsion. 
Here we explain the effect by comparing the system level structure calculated using two-level approximation (TLA) and the full model (including two-photon states), as illustrated in Fig.~\ref{fig:TLA}.
The single-excitation manifolds ($\mathcal{M}_1$, cyan part) have the same level structure in the two cases.
The coupling between $\rm Q_1 (Q_{2})$ and the coupler causes the energy of $|100\rangle$ ($|001\rangle$) to shift relative to the corresponding diabatic state.
In the dispersive regime, the frequency shift is about $\delta_{\rm 1c} \simeq g_{\rm 1c}^2/\Delta_{\rm 1c}$ ($\delta_{\rm 2c} \simeq g_{\rm 2c}^2/\Delta_{\rm 2c}$).
Since energy shifts ($\pm \delta_{12}$) caused by direct coupling between the qubits cancel each other, the total frequency shift of $|100\rangle$ and $|001\rangle$ is $\delta_{\rm 1c}+\delta_{\rm 2c}$.
The double-excitation manifold ($\mathcal{M}_2$, yellow part) has three diabatic states $|101\rangle^0$, $|011\rangle^0$ and $|110\rangle^0$ in the TLA case. The energy shift of $|101\rangle$ is $\delta_{\rm 1c}+\delta_{\rm 2c}$, which is the summation of the repulsion caused by the coupling between the qubits and the coupler.
Therefore, the ZZ interaction strength $\zeta$ is zero or negligible (very small ZZ interaction arises from the weak direct coupling between $\rm Q_1$ and $\rm Q_2$).
While in the full model case, $\mathcal{M}_2$ is complicated by three extra states $|002\rangle^0$, $|020\rangle^0$ and $|200\rangle^0$. 
Those states cause extra repulsion on $|101\rangle^0$ through direct or indirect interactions, resulting in non-zero $\zeta$. 
Strong ZZ interaction emerges when the $\mathcal{M}_2$ becomes more compact.

\onecolumngrid

\setcounter{equation}{0}
\renewcommand{\theequation}{B\arabic{equation}}	

\section*{APPENDIX B: Perturbation calculation of $\zeta$}

The analytical form of $\zeta$ in the dispersive regime is derived using fourth-order perturbation theory (with $\Delta_{ij}=\omega_{i}-\omega_{j}$, $\Sigma_{ij}=\omega_{i}+\omega_{j}$),
\onecolumngrid
\begin{eqnarray}
\begin{aligned}
\zeta_p=&\zeta_1+\zeta_2+\zeta_3+\zeta_4,  \\
\zeta_1=&0, \\
\zeta_2=&-\frac{2g_{12}^2}{\Delta_{12}+\alpha_1}+\frac{2g_{12}^2}{\Delta_{12}-\alpha_2}+\frac{2g_{12}^2}{\Sigma_{12}+\alpha_1}+\frac{2g_{12}^2}{\Sigma_{12}+\alpha_2}-\frac{4g_{12}^2}{\Sigma_{12}+\alpha_2+\alpha_1}, \\
\zeta_3=&2g_{\rm 1c}g_{12}g_{\rm 2c}
[ \frac{1}{\Delta_{12}-\alpha_2}(\frac{2}{\Delta_{\rm 1c}}
-\frac{2}{\Sigma_{\rm 2c}+\alpha_2})
- \frac{1}{\Delta_{12}+\alpha_1}(\frac{2}{\Delta_{\rm 2c}}
-\frac{2}{\Sigma_{\rm 1c}+\alpha_1}) + \frac{1}{\Delta_{12}}(-\frac{1}{\Delta_{\rm 1c}}  + \frac{1}{\Delta_{\rm 2c}} -
\frac{1}{\Sigma_{\rm 1c}}+\frac{1}{\Sigma_{\rm 2c}}) \\
&+\frac{1}{\Delta_{\rm 1c}\Delta_{\rm 2c}}+\frac{1}{\Delta_{\rm 1c}\Sigma_{12}}+\frac{1}{\Delta_{\rm 2c}\Sigma_{12}}-\frac{2}{\Delta_{\rm 2c}(\Sigma_{\rm 1c}+\alpha_1)}-\frac{4}{(\Sigma_{12}+\alpha_1)(\Sigma_{\rm 1c}+\alpha_1)}+\frac{4}{(\Sigma_{\rm 1c}+\alpha_1)(\Sigma_{12}+\alpha_1+\alpha_2)} + \frac{1}{\Delta_{\rm 1c}\Sigma_{\rm 2c}}\\
&+\frac{1}{\Delta_{\rm 2c}\Sigma_{\rm 1c}}-\frac{2}{\Sigma_{\rm 1c}(\Sigma_{12}+\alpha_2) } + \frac{1}{\Sigma_{12}\Sigma_{\rm 1c}}  - \frac{2}{\Delta_{\rm 1c}(\Sigma_{\rm2c}+\alpha_2)} +\frac{2}{(\Sigma_{12}+\alpha_2)(\Sigma_{\rm 2c}+\alpha_2)} + \frac{4}{(\Sigma_{12}+\alpha_1+\alpha_2)(\Sigma_{\rm 2c}+\alpha_2)} \\
&+ \frac{4}{(\Sigma_{12}+\alpha_1+\alpha_2)(\Sigma_{\rm 2c}+\alpha_2)} - \frac{2}{(\Sigma_{\rm 2c}+\alpha_2)\Sigma_{\rm 1c} } -\frac{2}{(\Sigma_{12}+\alpha_1)\Sigma_{\rm 2c}}+\frac{1}{\Sigma_{12}\Sigma_{\rm 2c}}+\frac{1}{\Sigma_{\rm 1c}\Sigma_{\rm 2c}}-\frac{2}{(\Sigma_{\rm 1c}+\alpha_1)\Sigma_{\rm 2c}}
], \\
\zeta_4 = &g_{\rm 1c}^2g_{\rm 2c}^2 [\frac{1}{\Delta_{12}}
( (\frac{1}{\Delta_{\rm 2c}}-\frac{1}{\Sigma_{\rm 1c}} )^2
-(\frac{1}{\Delta_{\rm 1c}}-\frac{1}{\Sigma_{\rm 2c}})^2 ) 
+\frac{2}{\Delta_{12}-\alpha_2}(\frac{1}{\Delta_{\rm 1c}}-\frac{1}{\Sigma_{\rm 2c}+\alpha_2})^2
-\frac{2}{\Delta_{12}+\alpha_1}(\frac{1}{\Delta_{\rm 2c}}-\frac{1}{\Sigma_{\rm 1c}+\alpha_1})^2
\\  
&-\frac{1}{\Delta_{\rm 1c}\Delta_{\rm 2c}}(\frac{1}{\Delta_{\rm 1c}+\Delta_{\rm 2c}}) 
+\frac{2}{\Delta_{\rm 1c}+\Delta_{\rm 2c}-\alpha_{\rm c}}(\frac{1}{\Delta_{\rm 1c}}+\frac{1}{\Delta_{\rm 2c}})^2
+\frac{1}{\Sigma_{12}}(\frac{1}{\Delta_{\rm 1c}}+\frac{1}{\Delta_{\rm 2c}})^2
+\frac{1}{\Sigma_{\rm 1c}\Delta_{\rm 2c}^2} + \frac{1}{\Sigma_{\rm 2c}\Delta_{\rm 1c}^2}-\frac{2}{\Delta_{\rm 1c}\Delta_{\rm 2c}\omega_{\rm}} + ...]\\
&+g_{\rm 1c}^2g_{12}^2
[\frac{1}{\Delta_{\rm 1c}\Delta_{12}^2} 
-\frac{1}{\Delta_{\rm 2c}\Delta_{12}^2} 
-\frac{2}{\Delta_{\rm 1c}(\Delta_{12}+\alpha_1)^2} -\frac{2}{\Delta_{\rm 1c}(\Delta_{12}-\alpha_2 )^2} 
+\frac{4}{\Delta_{\rm 2c}(\Delta_{12}+\alpha_1 )^2}+ ...  ] \\
&+g_{\rm 2c}^2g_{12}^2 
[\frac{1}{\Delta_{\rm 2c}\Delta_{12}^2} 
-\frac{1}{\Delta_{\rm 1c}\Delta_{12}^2}   -\frac{2}{\Delta_{\rm 2c}(\Delta_{12}+\alpha_1)^2}  -\frac{2}{\Delta_{\rm 2c}(\Delta_{12}-\alpha_2)^2}  +\frac{4}{\Delta_{\rm 1c}(\Delta_{12}-\alpha_2)^2}+ ...] \\
&+4g_{12}^4[ \frac{1}{(\Delta_{12}+\alpha_1)^3} -\frac{1}{(\Delta_{12}-\alpha_2)^3} + \frac{1}{(\Delta_{12}+\alpha_1)(\Delta_{12}-\alpha_2)^2} 
-\frac{1}{(\Delta_{12}+\alpha_1)^2(\Delta_{12}-\alpha_2)} + ...]. \\
\label{eq:zeta_pertur}
\end{aligned}
\end{eqnarray}

\begin{figure}
	\includegraphics[width=8.5cm]{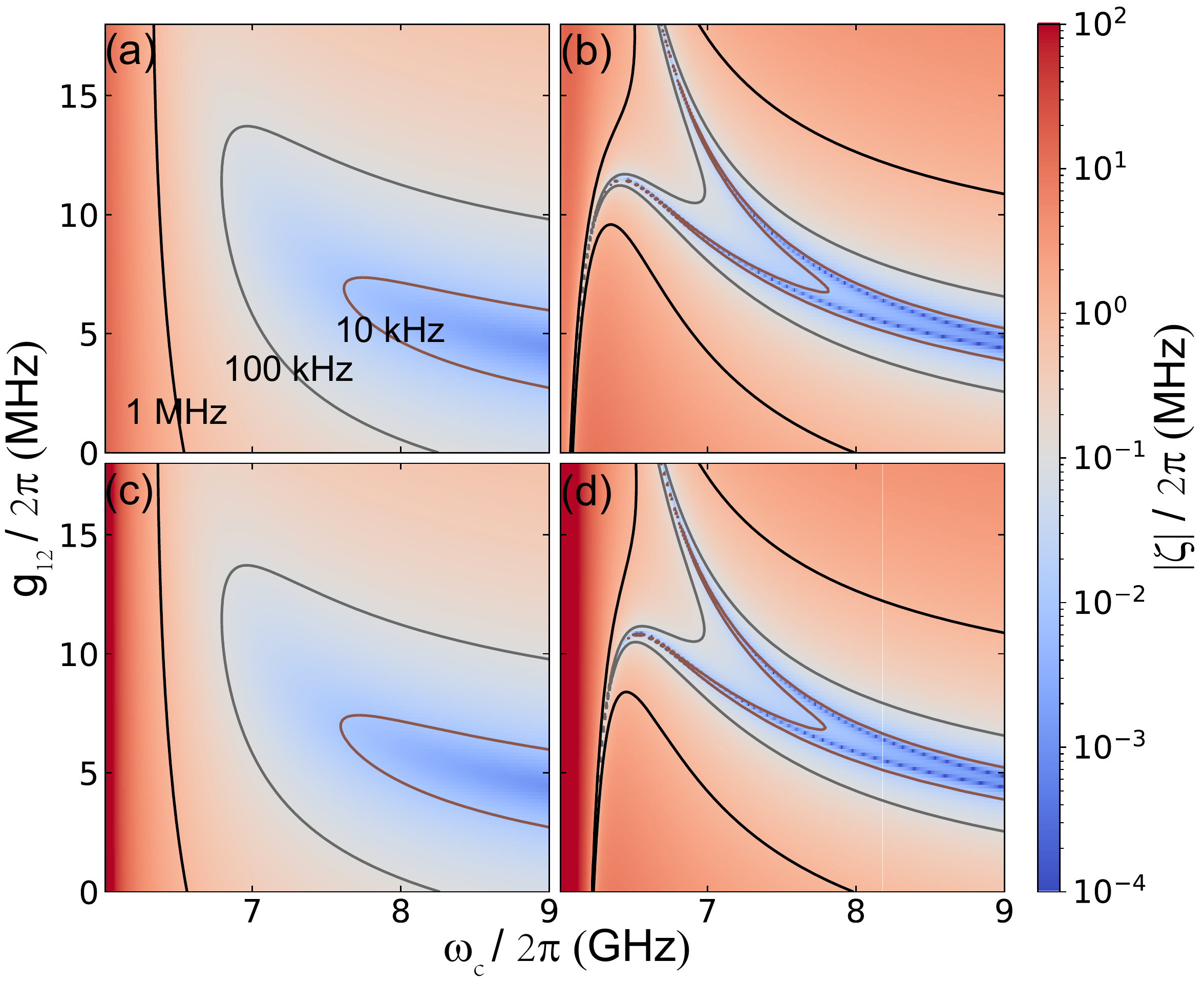}
	\caption{			
		Absolute ZZ interaction strength versus $\omega_c$ and $g_{12}$, calculated using numerical diagonalization (a,b) and Eq.~(\ref{eq:zeta_quad}) (c,d).
		The detuning between the qubits ($\Delta_{12}/2\pi$) is set at 600 MHz for the large-detuning case (a,c) and 150 MHz for the small-detuning case (b,d).  
		The other parameters are: $\alpha_{1,2,\rm c}/2\pi=-250, -250, -300$ MHz and $g_{\rm 1c, \rm 2c}/2\pi=120, 100$ MHz.
		\label{fig:ZZ_compare}}
\end{figure}

Here $\zeta_i$ denotes the $i$-th order perturbation result.
With assumptions: $\Sigma_{i \rm c}>>\Delta_{i \rm c}>>\Delta_{12} \sim g_{i\rm c} >> g_{12}$ ($i=1,2$), $\zeta$ can be approximately expressed as Eq.~(\ref{eq:zeta_quad}) in Sec.\Rmnum{2}.C.
In Fig.~\ref{fig:ZZ_compare}, we show that Eq.~(\ref{eq:zeta_quad}) is a good approximation by comparing the results with numerical diagonalization.

\setcounter{equation}{0}
\renewcommand{\theequation}{C\arabic{equation}}	

\section*{Appendix C: Noises in two-qubit RB }

Here we research noise contribution to gate errors measured by interleaved randomized benchmarking (RB). 
A standard interleaved-RB has the following steps:
1. apply a sequence control of $m$ random Cliffords to the ground state system;
2. append a final recovery Clifford to terminate the system at a target state $|\psi_{\rm th}\rangle$ (usually the ground state); 
3. average sequence fidelity over $k$ different sequences. The sequence fidelity is evaluated by compared the final state $|\psi_{\rm f}\rangle$ to the theoretical one, as $F_{\rm ref} = |\langle \psi_{\rm f}| \psi_{\rm th} \rangle |^2$;
4. interleave a specific gate within the random Cliffords, and repeat step.2 and step.3 to get the sequence fidelity $F_{\rm gate}$;
5. get the decay rate of the sequences $p_{\rm ref},p_{\rm gate}$ by fitting $F_{\rm ref} = Ap_{\rm ref}^m+B$ and $F_{\rm gate } = Ap_{\rm gate}^m+B$. The gate error is evaluated as $r_{\rm gate} = (1-p_{\rm gate}/p_{\rm ref}) (d-1)/d$, where $d$ is the dimension of the computational subsystem.

The Clifford group for two-qubit RB has a full size of 11520~\cite{barends2014superconducting}. The control sequence generates 60 different states with equal probability in theory, as shown in Table.~1.
In case of small errors, the instantaneous state during two-qubit RB sequence is close to one the 60 states.
Therefore, the error of the interleaved gate can evaluated by averaging 60 initial states, as
\begin{equation}
\epsilon =\frac{1}{60} \sum_{k=1,2,..,60} 1-\rm{Tr}(P (\rho_k) \rho_k), 
\label{eq:rb_error}
\end{equation}
where $P$ is the super-operator related to the gate and $\rho_k= |\psi_k\rangle \langle \psi_k |$ is the $k$-th initial state.

\subsection*{Transition errors }

\begin{figure}
	\includegraphics[width=8.5cm]{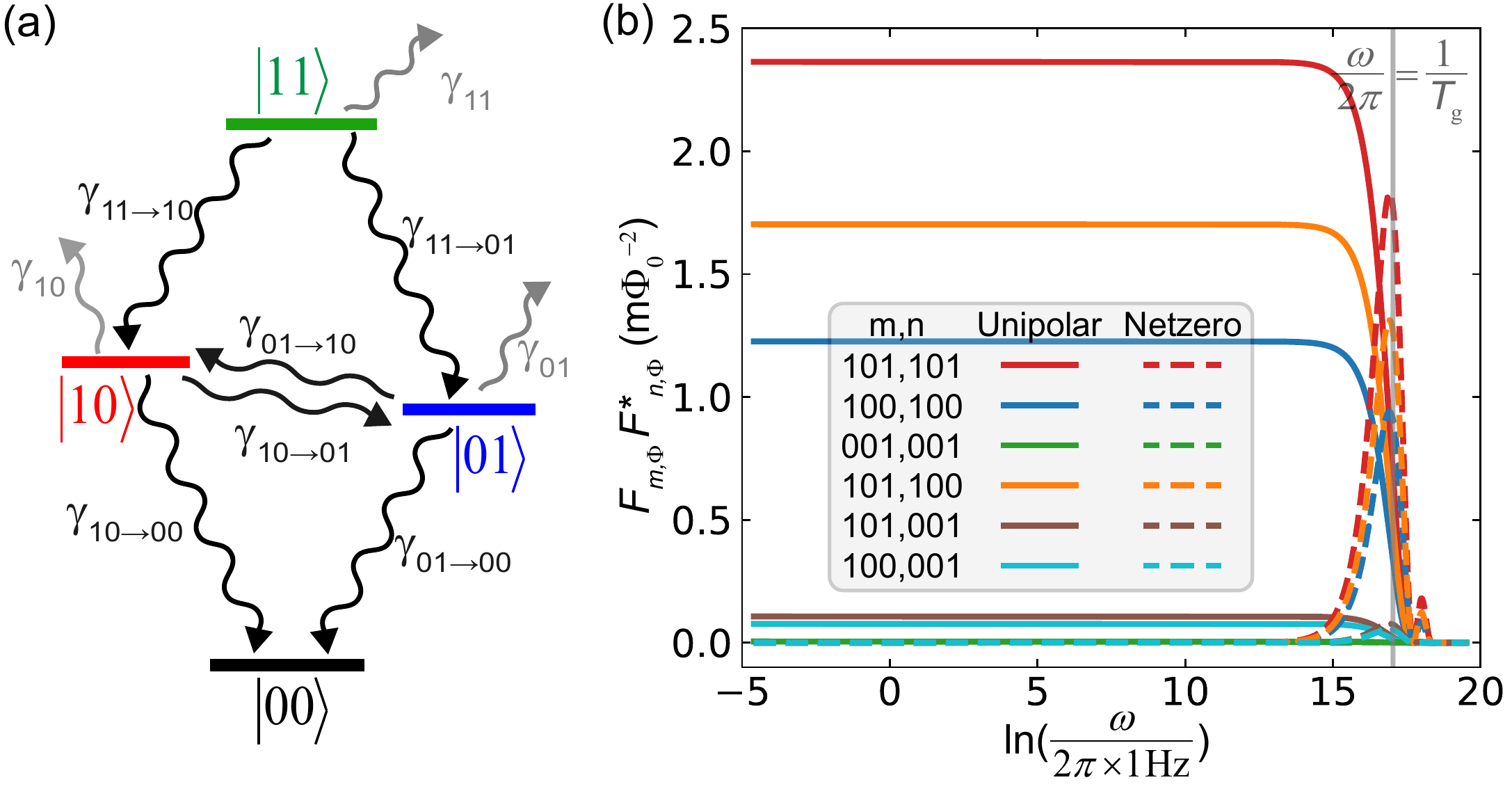}
	\caption{ 	
		(a) Sketch of noise-induced transitions. The black and gray wavy lines represent transitions within the computational subspace ($\mathcal{S} \to \mathcal{S}$) and leakage ($\mathcal{S} \to \mathcal{L}$). 
		(b) Phase covariance sensitivities to flux noise on the coupler at different frequencies, in a 40~ns unipolar (solid) and Net-Zero (dashed) CZ gate.
		The circuit parameters are the same as the ones used in Fig.~\ref{fig:FIG 7}.
		The AWP method is used in generating both the unipolar and NetZero pulse, with the coupler idly biased at 7.87 GHz (minimum residual ZZ) and 8.2 GHz (the sweet spot) respectively.
		\label{fig:noise_sketch}}
\end{figure}

The noise-induced transition errors in a two-qubit system are sketched in Fig.~\ref{fig:noise_sketch}, including: intra-$\mathcal{S}$ transitions ($\gamma_{01 \rightarrow 00},\gamma_{10 \rightarrow 00},\gamma_{11 \rightarrow 01},\gamma_{11 \rightarrow 10},\gamma_{01 \rightarrow 10},\gamma_{10 \rightarrow 01}$) and state leakage to the leakage subspace $\mathcal{L}$. 
The inverse processes including the noise-induced excitation and the transitions from $\mathcal{L}$ to $\mathcal{S}$ are ignored.
Since the transitions are related to noises at the energy gap between corresponding states ($\sim$ GHz), the noise correlation time ($\sim$ 1 ns) is much shorter than a typical gate time ($\gtrsim$ 10 ns). 
Therefore, the noise-induced transitions are treated as Markovian processes, which can be modeled using the Lindblad equation
\begin{equation}
\dot{\rho}(t) = -i[H(t),\rho(t)]+ \sum_{i}(c_i(t)\rho(t)c_i^{\dagger}(t) - \frac{1}{2}\{c_i^{\dagger}(t)c(t),\rho(t)\} ),
\label{eq:master_equation}
\end{equation}
where $c_i(t)$ is the jump operator related to a transition.
Considering state leakage out of $\mathcal{S}$, the density matrix $\rho$ is defined with basis $\{|00\rangle,|01\rangle,|10\rangle,|11\rangle,|l \rangle\}$. Here, $|l\rangle$ represents a state in $\mathcal{L}$.
As an example, the jump operator for the leakage of $|11\rangle$ (to $|l\rangle$) is
\begin{equation}
c_{11} = \sqrt{\gamma_{11}}\begin{pmatrix}
0 & 0 & 0 & 0 & 0 \\
0 & 0 & 0 & 0 & 0 \\
0 & 0 & 0 & 0 & 0 \\
0 & 0 & 0 & 0 & 0 \\
0 & 0 & 0 & 1 & 0 \\
\end{pmatrix},
\label{eq:leak_matrix}
\end{equation}
where $\gamma_{11}$ represents the leakage rate. 
For an initial state $(|01\rangle+|10\rangle+|01\rangle+|11\rangle)/2$ and $I$ gate, the error rate caused by the leakage is $\bar{\gamma}_{11} \tau/4$, where $\bar{\gamma}_{11}$ represents the average leakage rate during the gate and $\tau$ is the gate time.
We show transition errors for each of the 60 states in Table.~1.
Since the transition errors are unrelated to the phases of the states, the errors of a phase-type gate are the same as the ones of $I$ gate.
Therefore, the average transition error for a phase-type gate during RB is
\begin{equation}
\begin{aligned}
\epsilon_{\rm tr} &= \frac{1}{4}(\bar{\gamma}_{11}+\bar{\gamma}_{01}+\bar{\gamma}_{10}) \tau
 +\frac{1}{5}(\bar{\gamma}_{10\rightarrow 01} +\bar{\gamma}_{01\rightarrow 10} +\bar{\gamma}_{11\rightarrow 10}+\bar{\gamma}_{11\rightarrow 01}+\bar{\gamma}_{10 \rightarrow 00} +\bar{\gamma}_{01\rightarrow 00} )\tau
\end{aligned}
\label{eq:leak_error}
\end{equation}

\subsection*{Dephasing errors}

Pure dephasing errors are related to the frequency fluctuation of the computational states.
Assume that the theoretical state after a specific gate is  
\begin{equation}
\begin{aligned}
\Psi_{\rm th} &=\frac{|00\rangle}{2}+\frac{|01\rangle}{2}+\frac{|10\rangle}{2}+\frac{|11\rangle}{2}.
\label{eq:psi_state_ideal}
\end{aligned}
\end{equation}
Due to the fluctuation, the actual state becomes 
\begin{equation}
\begin{aligned}
\Psi &=\frac{|00\rangle}{2}+e^{i\Delta \phi_{01}}\frac{|01\rangle}{2}+e^{i\Delta \phi_{10}}\frac{|10\rangle}{2}+e^{i\Delta \phi_{11}}\frac{|11\rangle}{2},
\label{eq:psi_state}
\end{aligned}
\end{equation}
where $\Delta \phi_{m} = \int_{0}^{\tau} \delta {\widetilde{\omega}_m}(t) dt $ ($m = 01,10,11$) is the erroneous phase, and $\delta {\widetilde{\omega}_m}(t)$ is the frequency fluctuation of eigenstate $|m\rangle$.
The corresponding phase error is
\begin{equation}
\begin{aligned}
\epsilon_{\phi} & = 1- |\langle \Psi | \Psi_{\rm th} \rangle |^2 
\approx \frac{3}{16}\sum_{m=01,10,11} {(\Delta \phi_{m})}^2 - \frac{1}{8}[\Delta \phi_{10}\Delta \phi_{01}+\Delta \phi_{10}\Delta \phi_{11}+\Delta \phi_{01} \Delta \phi_{11}] ,
\label{eq:psi_error}
\end{aligned}
\end{equation}
where the approximation is established at small erroneous phases.

There are also 60 theoretical states with equal probability. Corresponding errors are shown in Table.~1. The average error caused by dephasing during RB is
\begin{equation}
\begin{aligned}
\epsilon_{\rm th} =&\frac{1}{20} [3 \langle (\Delta \phi_{10})^2 \rangle 
+3 \langle (\Delta \phi_{01})^2 \rangle 
+3 \langle (\Delta \phi_{11})^2 \rangle  
-2 \langle \Delta \phi_{01} \Delta \phi_{10} \rangle
-2 \langle \Delta \phi_{01} \Delta \phi_{11} \rangle
-2 \langle \Delta \phi_{10} \Delta \phi_{11} \rangle]. 
\end{aligned}
\label{eq:dephasing_error}
\end{equation}
The equation holds regardless of the noise model.

We next evaluate the phase covariance $\langle \phimn{m} \phimn{n}\rangle$. A series of independent noises (with a mean value of zero) $x_i(t),i=1,2,3...$ are assumed. The noise-induced phase accumulation on eigenstate $|m\rangle$ is
\begin{equation}
\phimn{m}  = \sum_i \int_{0}^{\tg} \pwpxt{m}{i} \dxt{i} \ud t,
\end{equation}
where $\pwpxt{m}{i}$ is the first-order frequency sensitivity of eigenstate $|m\rangle$ to the $i$-th noise (higher order terms are ignored in case of weak noise).
Then 
\begin{equation}
\begin{aligned}
\langle \phimn{m} \phimn{n}  \rangle &=  \langle \sum_i  \sum_j \int_{0}^{\tg} \pwpx{m}{i}{1} \dx{i}{1} \ud t_1  \int_{0}^{\tg} \pwpx{n}{j}{2} \dx{j}{2} \ud t_2  \rangle \\
&=   \sum_i  \sum_j \int_{0}^{\tg} \pwpx{m}{i}{1} \ud t_1  \int_{0}^{\tg} \pwpx{n}{j}{2}  \ud t_2  \langle \dx{i}{1} \dx{j}{2}  \rangle  \\
& =  \sum_i  \int_{0}^{\tg} \pwpx{m}{i}{1}  \ud t_1  \int_{0}^{\tg} \pwpx{n}{i}{2}  \ud t_2  \langle \dx{i}{1} \dx{i}{2}  \rangle  ,
\end{aligned}
\end{equation}
The last equation holds because the correlation function between two independent source is zero, as $\langle \dx{i}{1} \dx{j}{2}  \rangle = 0$.
Since the correlation function of a signal is the Fourier transformation of it's power spectral density (PSD) $S_i(\omega)$, we further get
\begin{equation}
\begin{aligned}
\langle \phimn{m} \phimn{n}  \rangle &= \sum_i  \int_{0}^{\tg} \pwpx{m}{i}{1}  \ud t_1  \int_{0}^{\tg}  \pwpx{n}{i}{2} \ud t_2  \frac{1}{2\pi} \int_{-\infty}^{\infty}  \ud \omega S_i(\omega) e^{-i\omega (t_2-t_1)} \\
& =  \sum_i \frac{1}{2\pi} \int_{-\infty}^{\infty}  \ud \omega S_i(\omega)  \int_{0}^{\tg}  e^{i\omega t_1} \pwpx{m}{i}{1} \ud t_1 \int_{0}^{\tg} e^{-i\omega t_2} \pwpx{n}{i}{2}   \ud t_2       \\
& = \sum_i \frac{1}{2\pi} \int_{-\infty}^{\infty} S_i(\omega)  F_{m,i}(\omega)F^*_{n,i}(\omega) \ud \omega,
\end{aligned}
\end{equation}
where $F_{m,i}(\omega)$ is the PSD of function
\begin{equation}
P_m(x_i,\tg;t) = \begin{cases}
\pwpxt{m}{i} & 0 \leq t \leq \tg\\
0& \rm{else},
\end{cases}
\end{equation}
And $F_{m,i}(\omega)F^*_{n,i}(\omega)$ is the covariance sensitivity to noise $x_i$ at frequency $\omega$. 
As an example, we show phase error sensitivities of CZ gates (40~ns) to flux noises on the coupler in Fig.~\ref{fig:noise_sketch}(b), based on unipolar and Net-Zero(NZ) modulation~\cite{rol2019fast}. 
The NZ pulse consists of two 20~ns unipolar pulses with opposite amplitude, making the gate fidelity insensitive to low-frequency flux noises.

~\\
\textbf{Long-range-correlated noise: quasistatic noise}

The PSD of quasistatic noise is $S_i(\omega) = 2\pi \sigma_i^2 \delta(\omega)$. The phase covariance is
\begin{equation}
\begin{aligned}
\langle \phimn{m} \phimn{n}  \rangle &= \sum_i (\sigma_i  \int_{0}^{\tg} \pwpx{m}{i}{1}  \ud t_1)  (\sigma_i  \int_{0}^{\tg} \pwpx{n}{i}{2} \ud t_2)  = \sum_i 2   \int_{0}^{\tg} \gammacorr{m}{i}(t) \ud t \int_{0}^{\tg} \gammacorr{n}{i}(t)  \ud t .
\end{aligned}
\end{equation}
where $ \gammacorr{m}{i}(t) = \frac{\sigma_i}{\sqrt{2}}  \pwpxt{m}{i} $ is the self-dephasing rate of eigenstate $|m\rangle$ caused by the quasistatic noise $x_i$. 

~\\
\textbf{$1/f$ noise}

The PSD of $1/f$ noise is $S_i(\omega) = \frac{A_i}{|\omega|/2\pi} $ and the phase covariance is
\begin{equation}
\begin{aligned}
\langle  \phimn{m} \phimn{n}  \rangle 
& = \sum_i A_i \int_{-\infty}^{\infty} \frac{1}{|\omega|}   F_{m,i}(\omega)F^*_{n,i}(\omega) \ud \omega \\
& \overset{s = \ln \omega }{=} 2\sum_i A_i \int_{\ln \omega_{\rm ir} }^{\ln \omega_{\rm ir} }   F_{m,i}({\rm{e}}^s)F^*_{n,i}({\rm{e}}^s) \ud s, \\
\end{aligned}
\end{equation}
where $ \omega_{\rm ir} $ ($ \omega_{\rm uv} $) is the low-frequency (high-frequency) cutoff of the $1/f$ noise. 

Here we consider the $1/f$-flux noise on the coupler. 
As depicted in Fig.~\ref{fig:noise_sketch}, for unipolar flux modulation, the phase covariance mainly affected by low frequency noise and the sensitivity term $F_{m,\Phi}(\omega)F^*_{n,\Phi}(\omega) \approx F_{m,\Phi}(0)F^*_{n,\Phi}(0)$ with $\omega/2\pi \ll 1/\tg$.
Then,
\begin{equation}
\begin{aligned}
\langle  \phimn{m} \phimn{n}  \rangle 
&\approx 2 (\ln \omega_{\rm uv} - \ln \omega_{\rm ir}) F_{m,i}(0)F^*_{n,i}(0) \sum_i A_i \\
&=  2 \ln (\frac{ \omega_{\rm uv}}{ \omega_{\rm ir}}) \int_{0}^{\tg} \pwpx{m}{i}{1}  \ud t_1 \int_{0}^{\tg} \pwpx{n}{i}{2} \ud t_2 \sum_i A_i \\
& =  \sum_i 2   \int_{0}^{\tg} \gammaf{m}{i}(t) \ud t \int_{0}^{\tg} \gammaf{n}{i}(t)  \ud t .
\end{aligned}
\end{equation}
The $1/f$-flux noise can be approximated as a quasistatic noise for unipolar modulation, with $\gammaf{m}{i}(t) = \frac{\sigma_i}{\sqrt{2}}  \pwpxt{m}{i} $ and $\sigma_i^2 = 2 A_i  \ln (\frac{ \omega_{\rm uv}}{ \omega_{\rm ir}})$. A typical choice for the cut-off frequencies is the $ \omega_{\rm uv}/2\pi = 1/T_{{\rm experi}} \sim 0.01$~Hz and $ \omega_{\rm ir}/2\pi \approx 1/T_{{\rm g}} \sim 10$~MHz.
While for the Net-Zero pulse, the low-frequency part is significantly suppressed and the gate mainly suffers from noise at a frequency close to $1/\tg$.

~\\
\textbf{Uncorrelated noise (white noise) }

For white noise $S_i(\omega) = A_i $, we can use the Parseval's theorem:
\begin{equation}
\int_{-\infty}^{\infty} P_i(t)P_j(t) dt = \frac{1}{2\pi} \int_{-\infty}^{\infty} F_i(\omega)F^{*}_j(\omega) d\omega.
\end{equation}
The phase covariance is then
\begin{equation}
\begin{aligned}
\langle \phimn{m} \phimn{n} \rangle &= \sum_i \frac{A_i}{2\pi} \int_{-\infty}^{\infty}  F_{m,i}(\omega)F^*_{n,i}(\omega) \ud \omega\\
&= \sum_i A_i \int_{-\infty}^{\infty}   P_m(x_i,\tg;t)  P_n(x_i,\tg;t)  \rm{d} t\\
&= \sum_i A_i \int_{0}^{\tg} \pwpxt{m}{i} \pwpxt{n}{i}  \rm{d} t\\
&  =\sum_i 2  \int_{0}^{\tg} \gammawhite{{mn}}{i}(t) \ud t.
\end{aligned}
\end{equation}
where
$\gammawhite{{mn}}{i}(t) = \frac{A}{2} \pwpxt{m}{i} \pwpxt{n}{i} $.
And $\gammawhite{{m(m)}}{i}(t) = \frac{A}{2}  (\pwpxt{m}{i})^2 $ is the self-dephasing rate of eigenstate $|m\rangle$ caused by white noise $x_i$.

\begin{table}[h] 
	\begin{tabular}{c|c|c|c|c} %设置了每一列的宽度，强制转换。	
		\hline
		\hline
		Number & States & Dephasing error & Leakage error & Intra-space transition error\\ %用&来分隔单元格的内容 \\表示进入下一行
		\hline %画一个横线，下面的就都是一样了，这里一共有4行内容
		&$|00\rangle$  & 0  & 0  & 0 \\
		4&$|01\rangle$  & 0  & $\bar{\gamma}_{01}\tau$  & $\bar{\gamma}_{01 \rightarrow 00}\tau$+$\bar{\gamma}_{01 \rightarrow 10}\tau$\\
		&$|10\rangle$  & 0  & $\bar{\gamma}_{10}\tau$  &  $\bar{\gamma}_{10 \rightarrow 00}\tau$+$\bar{\gamma}_{10 \rightarrow 01}\tau$ \\
		&$|11\rangle$  & 0  & $\bar{\gamma}_{11}\tau$  & $\bar{\gamma}_{11 \rightarrow 10}\tau$+$\bar{\gamma}_{11 \rightarrow 01}\tau$ \\
		\hline
		&($|00\rangle$ $\pm $ $(i) |01\rangle$)/$\sqrt{2}$ & $\langle (\Delta \phi_{01})^2 \rangle /4$  & $\gamma_{01}\tau$/2  & $\bar{\gamma}_{01 \rightarrow 00}\tau$/4+$\bar{\gamma}_{01 \rightarrow 10}\tau$/2 \\
		& & & & \\
		&($|00\rangle$ $\pm $ $(i) |10\rangle$)/$\sqrt{2}$ & $\langle (\Delta \phi_{10})^2 \rangle /4$    
		& $\bar{\gamma}_{10}\tau$/2  & $\bar{\gamma}_{10 \rightarrow 00}\tau$/4+$\bar{\gamma}_{10 \rightarrow 01}\tau$/2\\
		& & & & \\
		&($|00\rangle$ $\pm $ $(i) |11\rangle$)/$\sqrt{2}$ & $\langle (\Delta \phi_{11})^2 \rangle /4$   &  $\bar{\gamma}_{11}\tau$/2 & $\bar{\gamma}_{11 \rightarrow 10}\tau/2$+$\bar{\gamma}_{11 \rightarrow 01}\tau$/2 \\
		& & & & \\
		24&($|01\rangle$ $\pm $ $(i) |10\rangle$)/$\sqrt{2}$ & $\langle (\Delta \phi_{01}-\Delta \phi_{10})^2 \rangle /4$   & ($\bar{\gamma}_{01}\tau$+ $\bar{\gamma}_{10}\tau$)/2  & \makecell[c]{$\bar{\gamma}_{10 \rightarrow 00}\tau$/2+ $\bar{\gamma}_{01 \rightarrow 00}\tau$/2 \\+$\bar{\gamma}_{10 \rightarrow 01}\tau$/4+$\bar{\gamma}_{01 \rightarrow 10}\tau$/4}	\\
		& & & & \\
		&($|01\rangle$ $\pm $ $(i) |11\rangle$)/$\sqrt{2}$ & $\langle (\Delta \phi_{11}-\Delta \phi_{01})^2 \rangle /4$   & ($\bar{\gamma}_{01}\tau$+ $\bar{\gamma}_{11}\tau$)/2  & \makecell[c]{$\bar{\gamma}_{11 \rightarrow 10}\tau/2$+$\bar{\gamma}_{11 \rightarrow 01}\tau$/4 \\ +$\bar{\gamma}_{01 \rightarrow 00}\tau/2$+$\bar{\gamma}_{01 \rightarrow 10}\tau/2$}  \\
		& & & & \\
		&($|10\rangle$ $\pm $ $(i) |11\rangle$)/$\sqrt{2}$ & $\langle (\Delta \phi_{11}-\Delta \phi_{10})^2 \rangle /4$   & ($\bar{\gamma}_{10}\tau$+ $\bar{\gamma}_{11}\tau$)/2  & \makecell[c]{$\bar{\gamma}_{11 \rightarrow 10}\tau/4$+$\bar{\gamma}_{11 \rightarrow 01}\tau$/2 \\ +$\bar{\gamma}_{10 \rightarrow 00}\tau/2$+$\bar{\gamma}_{10 \rightarrow 01}\tau/2$} \\
		\hline
		32 & \makecell[c]{($|00\rangle+a|01\rangle+b|10\rangle+c|11\rangle$)/2 \\
			$a = \pm, \pm i$ \\
			$b = \pm, \pm i$ \\
			$c = \pm ab$ }&  \makecell[l]{[3 $\langle (\Delta \phi_{10})^2 \rangle$ 
			+3 $\langle (\Delta \phi_{01})^2 \rangle$ \\
			+3 $\langle (\Delta \phi_{11})^2 \rangle$ 
			-2  $\langle \Delta \phi_{01} \Delta \phi_{10} \rangle$ \\
			-2  $\langle \Delta \phi_{01} \Delta \phi_{11} \rangle$
			-2  $\langle \Delta \phi_{10} \Delta \phi_{11} \rangle$]/16} & $(\bar{\gamma}_{11}+\bar{\gamma}_{10}+\bar{\gamma}_{01})\tau$/4  &  \makecell[l]{$3(\bar{\gamma}_{11 \rightarrow 10}\tau + \bar{\gamma}_{11 \rightarrow 10}\tau$ \\ + $\bar{\gamma}_{10 \rightarrow 00}\tau+\bar{\gamma}_{01 \rightarrow 00}\tau$ \\ + $\bar{\gamma}_{01 \rightarrow 10}\tau + \bar{\gamma}_{10 \rightarrow 01}\tau  )/16$} \\
		\hline
		Average &   &  \makecell[l]{[3 $\langle (\Delta \phi_{10})^2 \rangle$ 
			+3 $\langle (\Delta \phi_{01})^2 \rangle$ \\
			+3 $\langle (\Delta \phi_{11})^2 \rangle$ 
			-2  $\langle \Delta \phi_{01} \Delta \phi_{10} \rangle$ \\
			-2  $\langle \Delta \phi_{01} \Delta \phi_{11} \rangle$
			-2  $\langle \Delta \phi_{10} \Delta \phi_{11} \rangle$]/20}  & $(\bar{\gamma}_{11}+\bar{\gamma}_{01}+\bar{\gamma}_{10})\tau/4$  &  \makecell[l]{$(\bar{\gamma}_{10\rightarrow 01} +\bar{\gamma}_{01\rightarrow 10}$ \\ $+\bar{\gamma}_{11\rightarrow 10}+\bar{\gamma}_{11\rightarrow 01}$\\ $+\bar{\gamma}_{10 \rightarrow 00} +\bar{\gamma}_{01
				\rightarrow 00} )\tau/5$} \\
		\hline
		\hline
	\end{tabular}
	\label{table_rb}
	\caption{60 states that appear with the same probability during two-qubit randomized benchmarking. The transition errors are evaluated for $I$ gate or phase-type gates, with $\bar{\gamma}, \tau$ represent the average transition rate and the gate time.} 
\end{table}

~\\
\twocolumngrid

\bibliographystyle{apsrev4-1_new}
\normalem
\bibliography{reference}

\end{document}